\RequirePackage{lineno}
\RequirePackage{float}
\documentclass[aps,physrev,twocolumn,superscriptaddress,linenumbers]{revtex4}
\pdfoutput=1 

\usepackage{soul}
\usepackage{amssymb}
\usepackage{amsmath}
\usepackage{graphicx}
\usepackage{extarrows}
\usepackage{url}
\usepackage{varioref}
\usepackage[hidelinks]{hyperref}
\usepackage[capitalise,compress]{cleveref}
\usepackage{soul}
\setcitestyle{super}
\usepackage[normalem]{ulem}
\usepackage{bm}
\usepackage{multibib}
\usepackage{orcidlink}
\usepackage{xcolor}

\usepackage{fancyhdr}
\fancypagestyle{firstpage}{
  \fancyhf{}

  \fancyfoot[L]{%
    \footnotesize%
    $^*$ These authors contributed equally to this work. \\
    Corresponding address: arinjoy@umd.edu
  }
}

\newcommand{\JQI}{Joint Quantum Institute and Joint Center for Quantum Information and Computer Science, NIST and University of Maryland, College Park, MD 20742, USA\\
\textsuperscript{$\dagger$}Present address: QuEra Computing Inc., Boston, MA 02135, USA}
\newcommand{\MSU}{Department of Physics and Astronomy, Michigan State University, East Lansing, MI 48824, USA}
\newcommand{\FRIB}{Facility for Rare Isotope Beams, East Lansing, MI 48824, USA}
\newcommand{\JHU}{Johns Hopkins University Applied Physics Laboratory, Laurel, MD 20723, USA}
\newcommand{\RICE}{Department of Physics and Astronomy, Rice University, Houston, TX, USA}
\newcommand{\DUKE}{Duke Quantum Center, Departments of Electrical and Computer Engineering and Physics, Duke University, Durham, NC 27708, USA}
\newcommand{\AWS}{Joint Quantum Institute and Joint Center for Quantum Information and Computer Science, NIST and University of Maryland, College Park, MD 20742, USA\\
\textsuperscript{$\ddagger$}Present Address: AWS Center for Quantum Computing, Pasadena, CA 91125, USA.}

\newcommand{\ket}[1]{\left \lvert #1 \right \rangle}
\newcommand\mean[1]{\ensuremath{\left\langle#1\right\rangle}}

\begin{document}

\title{Non-equilibrium critical scaling and universality in a quantum simulator}

\author{Arinjoy De$^{*,\dagger}$~\orcidlink{0000-0001-9184-8434}}
\affiliation{\JQI}
\author{Patrick Cook$^*$}
\affiliation{\MSU}
\affiliation{\FRIB}
\author{Mostafa Ali}
\affiliation{\MSU}
\author{Kate Collins}
\affiliation{\JQI}
\author{William Morong $^\ddagger$}
\affiliation{\AWS}
\author{Daniel Paz}
\affiliation{\MSU}
\author{Paraj Titum}
\affiliation{\JQI}
\affiliation{\JHU}
\author{Guido Pagano}
\affiliation{\RICE}
\author{Alexey V. Gorshkov}
\affiliation{\JQI}
\author{Mohammad Maghrebi}
\affiliation{\MSU}
\author{Christopher Monroe~\orcidlink{0000-0003-0551-3713}}
\affiliation{\JQI}
\affiliation{\DUKE}

\begin{abstract}
    Universality and scaling laws are hallmarks of equilibrium phase transitions and critical phenomena. However, extending these concepts to non-equilibrium systems is an outstanding challenge. Despite recent progress in the study of dynamical phases, the universality classes and scaling laws for non-equilibrium phenomena are far less understood than those in equilibrium. In this work, using a trapped-ion quantum simulator with single-spin resolution, we investigate the non-equilibrium nature of critical fluctuations following a quantum quench to the critical point. We probe the scaling of spin fluctuations after a series of quenches to the critical Hamiltonian of a long-range Ising model. With systems of up to 50 spins, we show that the amplitude and timescale of the post-quench fluctuations scale with system size with distinct universal critical exponents, depending on the quench protocol. While a generic quench can lead to thermal critical behavior, we find that a second quench from one critical state to another (i.e.~a double quench) results in a new universal non-equilibrium behavior, identified by a set of critical exponents distinct from their equilibrium counterparts. Our results demonstrate the ability of quantum simulators to explore universal scaling beyond equilibrium.
\end{abstract} 
\maketitle
\thispagestyle{firstpage}

\newpage

\section*{Introduction}
In recent years, substantial theoretical~\cite{Heyl2013, Heyl2019, Marino2022, Li2022} and experimental~\cite{Jurcevic2017, Zhang2017DPT} progress has been achieved in understanding emergent behavior of isolated quantum systems out of equilibrium. In this context, non-equilibrium many-body systems can be investigated by measuring quantum dynamics after a quench~\cite{calabrese2011}, namely after a change of the Hamiltonian parameters that is much faster than the typical energy scales in the system---which is routinely performed in AMO (atomic, molecular, and optical) systems.

Although quench dynamics are extremely complex in general, one would expect that macroscopic observables, after a short time, become insensitive to the microscopic details~\cite{sachdev2011}. In particular, in the vicinity of a phase transition, the dynamics should give rise to universal critical behavior which leads to scale-invariant spatio-temporal correlations with universal exponents~\cite{Paraj2020}. Notably, we show that non-equilibrium critical phenomena can emerge in the context of quench dynamics in one-dimensional systems with long-range interactions. In general, universal non-equilibrium phenomena are relevant far beyond the scope of AMO and condensed matter physics, including chemistry, biology, and even sociology~\cite{Stankovski2017}. Examples ranging from glassy transitions seen in polymers, colloidal gels, and spin glasses~\cite{Li2021} to symmetry-breaking transitions in the Universe after the `Big Bang'~\cite{ZUREK1996} all exhibit non-equilibrium critical behavior.

The unprecedented degree of control over quantum systems in platforms such as trapped ions~\cite{Monroe2021,Blatt2012}, ultracold atoms~\cite{Bloch2015,Bernien2017}, nitrogen-vacancy centers~\cite{Choi2017}, superconducting circuits~\cite{Schoelkopf2004,kai2020} and others~\cite{Kimble2018,Katori2008,Haroche2001} have made it possible to probe fundamental questions about non-equilibrium many-body physics including prethermalization~\cite{Gring2012,Neyenhuis2017}, many-body localization~\cite{Jake2016,Choi2016}, discrete time crystals~\cite{Choi2017,Zhang2017DTC}, and dynamical phase transitions~\cite{Jurcevic2017,Zhang2017DPT}. For example, universal scaling around non-thermal fixed points has been observed with Bose-Einstein condensates that exhibit self-similar behavior; these observations are however not related to an underlying critical behavior~\cite{Eigen2018,Erne2018,Prufer2018,Baganto2022,Jae2024}. In contrast, the work reported here is fundamentally tied to the existence of a phase transition and extends the remarkably rich domain of critical phenomena in equilibrium to far-from-equilibrium dynamics.

\begin{figure*}[t]
\includegraphics{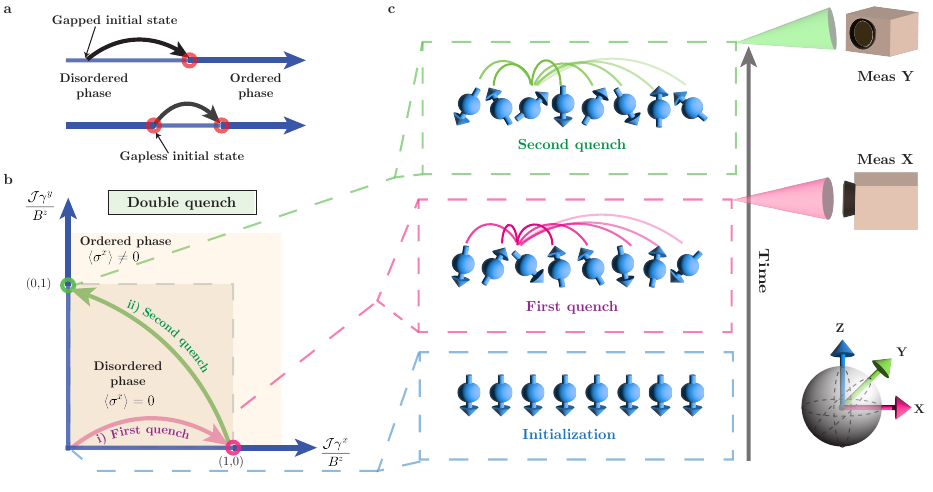}
\caption{\textbf{Critical quench dynamics. a.} Disorder-to-order phase transitions emerge even in the non-equilibrium setting of quench dynamics, and exhibit critical behavior at the transition. The ordered and disordered phases are shown in thick and thin lines, respectively, and the arrows indicate a quench to the critical point. While a quench from a gapped initial state (top panel) to the critical point (red circles) generically leads to effective thermal behavior, a quench from a gapless state (bottom panel), corresponding to a distinct critical point, gives rise to non-equilibrium criticality. 
\textbf{b.} Ground-state phase diagram with Ising interaction along $x$ or $y$ direction. ${\mathcal{J}\gamma^{x,y}/B^z}$ is the ratio of the Kac-normalized effective interaction strength ${\mathcal{J}\gamma^{x,y}}$ to the transverse field strength ${B^z}$ (see text). The phase boundary is shown in gray dashed lines, with red and green circles indicating the critical points where the quenches are performed. \textbf{c.} The experimental sequence starting with all spins initialized along ${\ket{\downarrow}_z}$. The first quench is applied with interactions along the ${x}$ direction, and the evolution is measured by projecting the spins along ${x}$. For the second quench, both the interaction and measurement bases are switched from ${x}$ to ${y}$ direction. In the double-quench experiment, the second quench is applied after evolving under the first quench, but no measurement is performed before the second quench. The curved lines illustrate the long-range interaction among all the spins where the opacity reflects interaction strengths that weaken
with distance.}\label{fig1} 
\end{figure*}

In this work, we investigate the critical behavior in the vicinity of a dynamical phase transition in quench dynamics. 
An immediate challenge is that a quantum quench generically leads to thermalization\cite{Polkovnikov_review_2011}, thus the resulting critical behavior is effectively thermal\cite{giamarchi2016strongly,Mitra_review_2018}. While this is the case in a quench from a gapped initial state, a quench from a gapless state (e.g., the critical point of the Hamiltonian) could lead to genuinely non-thermal critical behavior \cite{Cazalilla_2016,Chiocchetta_2017,Paraj2020,paul2022hidden} (see the schematic picture in~\cref{fig1}a). 
We attribute these features to the behavior of the \textit{soft mode}~\cite{Mitra_review_2018}. Specifically, we find that a generic quench from a gapped initial state excites this mode and heats it up, resulting in effective thermal behavior. In contrast, a critical-to-critical quench leads to \textit{over}heating of the \textit{soft mode}, giving rise to novel non-equilibrium critical behavior.

\section*{Results}
We probe the post-quench dynamics of a transverse-field Ising chain with tunable power-law interactions. The Hamiltonian of the model is represented as (${\hbar=1}$):

\begin{equation}
    H=-\sum_{i<j}^{N}J_{ij}(\gamma^x\sigma_i^x\sigma_j^x+\gamma^y\sigma_i^y\sigma_j^y)+B^z\sum_i^{N}\sigma_i^z ,
    \label{hamiltonian}
\end{equation}
where ${\sigma^{x,y,z}_i}$ are the Pauli matrices acting on the ${i}$'th spin. ${J_{ij}}$ is the interaction strength between ions ${i}$ and ${j}$, ${{B^z}}$ is the global transverse magnetic field and $N$ is the number of spins. The coefficients ${{(\gamma^x,\gamma^y)\in [0,1]}}$, and only one of them is non-zero during a single quench (Methods). The interaction strength falls off approximately following a power-law of the form ${J_{ij}\sim J/\lvert i-j\rvert^p}$, where ${J>0}$ is the effective interaction strength and ${p}$ represents the range of interaction~\cite{Monroe2021}. The exponent ${p}$ was tuned to be at 0.89 for all the experiments and all system sizes (Methods). We note that, for large system sizes, the interaction matrix elements deviate from a simple power-law behavior and exhibit faster decay at longer distances (see Supplementary Information (SI), Sec.~VII.B). In order to maintain a well-defined thermodynamic limit, here onwards, we refer to the interaction after it is Kac-normalized as ${\mathcal{J}=\frac{1}{N-1}\sum_{i,j}J_{ij}}$~\cite{kac1969}. We encode the quantum spins in the ground state hyperfine manifold of the ${^{171}}$Yb${^+}$ ions, where ${\ket{\downarrow}_z\equiv\ket{^{2}S_{1/2},F=0,m_F=0}}$ and ${\ket{\uparrow}_z\equiv\ket{^{2}S_{1/2},F=1,m_F=0}}$, and we perform high-fidelity state preparation and site-resolved detection using state-dependent fluorescence (Methods)~\cite{Olmschenk2007}.

The transverse-field Ising model exhibits a ground-state phase transition from a disordered paramagnet to a magnetically ordered state. As the ratio of Kac-normalized effective interaction field (${\mathcal{J}\gamma^x}$) to the transverse magnetic field (${B^z}$) is varied across the critical point ${\mathcal{J}\gamma^x/{B^z}=1}$, the average in-plane magnetization (${\mean{\sigma^x}}$) changes from zero (disordered phase) to a non-zero value (ordered phase) in a second-order phase transition (Fig.~\ref{fig1}a). By performing quenches to various values of ${\mathcal{J}\gamma^x/{B^z}}$, we identify the critical point of this phase transition and observe the characteristic divergent fluctuations. We report that, after a single quench from a gapped initial state, the critical behavior and exponents are effectively thermal; see \cref{fig1}a (top panel). In contrast, the scenario in \cref{fig1}a (bottom panel) involves preparing a critical initial state, which is experimentally challenging. Instead, we show that a sequence of quenches to multiple critical points (Fig.~\ref{fig1}b) achieves the same objective and leads to non-equilibrium critical behavior.

We begin with a single quench sequence where all the spins are initialized in the ${\ket{\downarrow\downarrow...\downarrow}_z}$ state, which is the gapped ground state of the initial Hamiltonian in the absence of the Ising interaction (SI Sec.~III). Then the spin system is evolved after an interaction quench of the form Eq.~\eqref{hamiltonian}, in which ${\gamma^x=1}$, ${\gamma^y=0}$ (Fig.~\ref{fig1}a). We measure the total spin ${S_x=\sum_i^N\sigma_i^x/2}$ projected along the direction of interaction (here along ${x}$) and calculate the net correlator defined as

\begin{equation}
    \mean{C^2_x}=\mean{S_x^2}-\mean{S_x}^2.
    \label{correlation}
\end{equation}
We characterize the dynamics through the net correlator since the Ising symmetry of the Hamiltonian 
and the initial state implies that the average magnetization along the ${x}$ direction remains zero at all times. The definition of ${\mean{C^2_x }}$ further removes any bias of the average magnetization due to imperfect single-qubit rotations in the experiment. In the Supplementary Fig. 1, we report the post-quench evolution of ${\mean {C_x^2 }/N^2}$ with 10 ions. The observed net correlator increases in amplitude and exhibits slower dynamics as ${B^z}$ is swept from larger to smaller values, consistent with numerical simulation that includes the decoherence effects of the experiment. This behavior hints at a continuous dynamical phase transition\cite{Marino2022}.

Equilibrium phase transitions are commonly identified by defining an order parameter that acquires a nonzero value as the system transitions from the disordered to the ordered phase.
Analogous to equilibrium phases, one may consider the in-plane magnetization to identify a dynamical phase transition. Using a mean-field analysis to compute the long-time average of the magnetization, we can identify ${B^z_c/\mathcal{J}=1}$ as the dynamical critical point of the disorder-to-order phase transition, which coincides with the ground-state critical point (SI Secs.~III \& V). While magnetization remains zero for our chosen initial state, we instead consider the temporal maximum of the net correlator, ${{\cal M}^2=\max_t\left [\mean{C^2_x}/N^2\right ]}$, as a proxy for the order parameter; the maximum is chosen to find a large signal in spite of decoherence.
\begin{figure}
\centering
\includegraphics{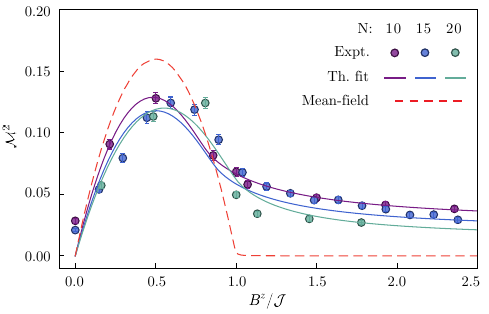}
\caption{\textbf{Phase transition from order parameter:} We report scaled maximum net correlator ${{\cal M}^2=\max_t\left [\mean{C^2_x }/N^2\right ]}$ as a function of ${B^z/\mathcal{J}}$ for system sizes ${N=10, 15, 20}$. The solid lines are obtained by fitting the experimental data to the finite size corrected order parameter (Eq.~(28) of SI), which has the critical point as a fit parameter. The extracted values are  ${0.83\:(19), 0.88\:(6), 1.01\:(9)}$ respectively for ${N=10, 15, 20}$; the difference from the predicted critical point ${B^z/\mathcal{J}=1}$ is due to finite-size effects and experimental imperfections. For simplicity, we use the predicted critical value for studies in Figs.~\ref{fig3}~and~\ref{fig4}. The dashed line represents the mean-field solution with an inflection point at ${B^z/{\cal J}=1}$. For the comparison of the experimental data against decoherence-free numerical simulation, see Supplementary Fig. 2. The error bars are statistical fluctuations around the mean value.
}\label{fig2}
\end{figure}
In Fig.~\ref{fig2}, we show ${{\cal M}^2}$ as a function of the scaled magnetic field strength ${B^z/\mathcal{J}}$. While there is no sharp transition for finite system sizes (${N=10, 15, 20}$), the order parameter clearly shows an inflection point around ${B^z/\mathcal{J}\sim1}$ and a peak at small ${B^z}$, indicating the onset of ordering. Notably, the observed order parameter qualitatively follows the mean-field prediction in the ordered phase, ${{\cal M}^2 \propto (B^z/\mathcal{J})(1-B^z/{\cal J})}$; see the dashed line in Fig.~\ref{fig2}. Moreover, one can even capture the finite-size corrections by considering fluctuations at finite system sizes. The solid lines in Fig.~\ref{fig2} depict the function describing the finite-size corrected order parameter, which has the critical point and an overall scale as fit parameters. The inferred critical values are well in agreement with the mean-field prediction (SI Secs.~IV.A \& IV.C). Having identified the dynamical critical point, the immediate questions are: What is the nature of the critical behavior at the phase transition, and does it genuinely go beyond the equilibrium paradigm?

\begin{figure*}
\includegraphics{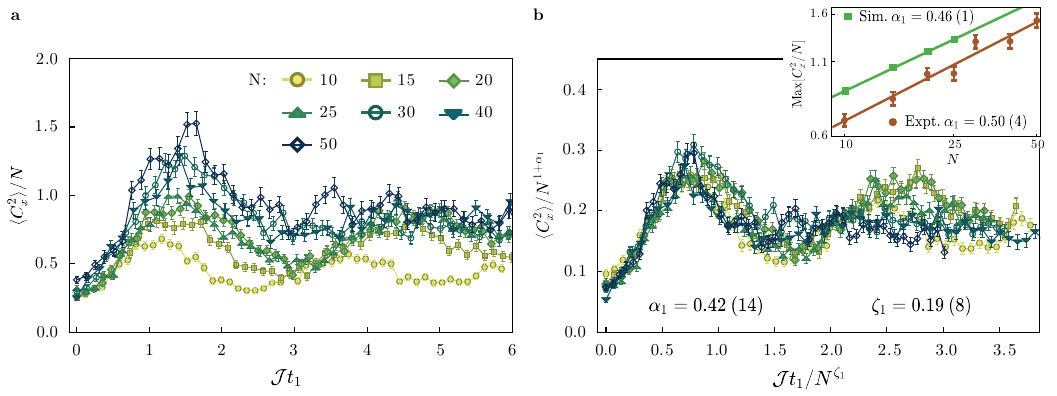}
\caption{\textbf{Unscaled (a) and scaled (b) fluctuations after a single quench.} \textbf{a,} We report experimental critical fluctuations with system sizes up to ${N=50}$ ions.
We obtain the critical scaling exponents (${\alpha_{1},\zeta_{1}}$) by optimizing the weighted Euclidean distance between each of the curves to get the best collapse for the experimental \textbf{b,} data [see main text for details]. We observe remarkable similarity between the exponents found in the experiment and simulations (see Supplementary Fig. 3), highlighting the universality of the exponents despite experimental imperfections as well as finite-size effects. Comparing the decoherence-free numerical simulation in Supplementary Fig. 3 with the experimental data, we note that the fluctuations in the experiment are reduced due to decoherence and imperfect detection fidelity; however, as we can see from the scaled data, the scaling collapse is still observed for all the system sizes. In the \textbf{inset b}, we find consistent scaling exponents by fitting the maximum values of the fluctuation (dots) to a power-law fit to ${N^{\alpha_1}}$ (solid lines). Although this method does not capture the full evolution, we get excellent agreement of exponents for both the simulation and the experiment. The data points from numerical simulation (in green) do not account for decoherence, resulting in higher peak values as compared to the experimental values (see Supplementary Fig. 3). In Supplementary Fig. 3 b,
we present numerical simulations of critical dynamics for ${N=10}$ and ${15}$,  incorporating decoherence effects. 
The error bars of the experimental data are statistical fluctuations around the mean value.} \label{fig3}
\end{figure*}
As a first step toward answering these questions, we experimentally scale up the single quench experiment to system sizes up to ${N=50}$ ions and observe the net correlator which, at the critical point, characterizes critical fluctuations. As we calibrate the quench Hamiltonian parameters to be at the (mean-field) critical point for all the system sizes, within our experimental uncertainty, we see that the fluctuations grow and evolve more slowly with increasing system size, indicating an emergent universal critical behavior (Fig.~\ref{fig3}a).
We identify similar behavior by a numerical simulation of the quench dynamics with experimental spin-spin coupling parameters for system sizes up to ${N=25}$ (Supplementary Fig. 3), the maximum number of spins that can be simulated using available computational resources. Such critical behavior leads to scaling relations which are independent of microscopic length/time scales~\cite{cardy1996scaling}. Using scaling analysis, we find the net correlator satisfies the functional form given by\cite{Paraj2020} 

\begin{equation}
    \mean{C_x^2}= N^{1+\alpha _1}f\left(\frac{\mathcal{J}t_1}{N^{\zeta_1}}\right),
    \label{scalefunction}
\end{equation}
where the exponent ${\alpha_1}$ characterizes the amplitude scaling of fluctuations with system size and ${\zeta_1}$ describes the dynamical scaling. We verify that the scaling relation and the experimentally-obtained exponents ${\alpha_1=0.42\:(14)}$, and ${\zeta_1 = 0.19\:(8)}$ (see Fig.~\ref{fig3}b), are consistent with the results of the exact simulation (see Supplementary Fig. 3). The procedure to determine exponents that yield the best collapse of the data is detailed in the SI Sec.~VIII. Remarkably, the above exponents are also consistent with mean-field exponents at the thermal phase transition~\cite{Paraj2020} ${\alpha_1=1/2,\;\zeta_1=1/4}$ (SI Sec.~IV.B). Additionally, we fit the maximum amplitude of fluctuations against ${N^{\alpha_1}}$ to obtain the exponent ${\alpha_1=0.50\:(4)}$ (see the inset of Fig.~\ref{fig3}b) which is in excellent agreement with that of thermal equilibrium. Indeed, it is expected that the latter procedure leads to a more precise exponent ${\alpha_1}$ since the dynamical features are more susceptible to decoherence. Note that the numerically obtained exponents do not account for decoherence due to the limitations of numerical methods for the system sizes considered here (Methods).

An effective thermal critical behavior is to be expected in quench dynamics \textit{if} a phase transition exists \cite{Mitra_review_2018,giamarchi2016strongly}, but this typically requires higher dimensions or, in this case, long-range interactions. The emergence of the thermal critical exponents does not mean that the system has thermalized. In fact, long-range interacting systems often exhibit prethermalization for a long window in time~\cite{Halimeh2017, Zunkovich2018, Neyenhuis2017}.
Instead, the thermal exponents should be attributed to the effective thermalization of the soft mode \cite{Mitra_review_2018}.
This can be understood through a Holstein-Primakoff transformation that maps spins to bosonic variables, ${\sigma^x_i \to \frac{1}{\sqrt{N}}(a_i + a_i^\dagger)}$, a mapping that is valid near a fully polarized state along the ${z}$ direction. The lowest energy excitation of the system corresponds to a collective excitation of a bosonic mode, characterized by the operator ${a\equiv \sum_i a_i}$, which becomes gapless (\textit{softens}) at the phase transition. The total spin can be then described as ${S_x \to \sqrt{N} x}$ where ${x \sim a+a^\dagger}$ may be interpreted as the position operator of a harmonic oscillator with a characteristic frequency ${\Omega}$ which vanishes at the phase transition.

\begin{figure*}
\includegraphics{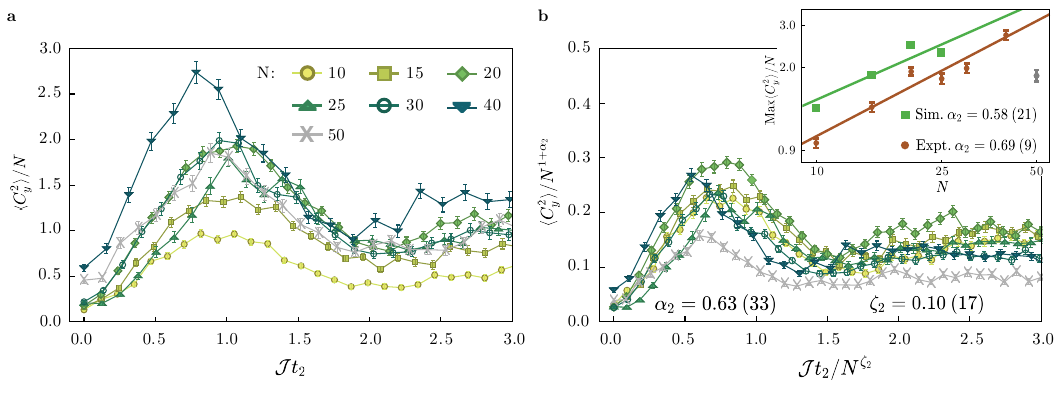}
\caption{\textbf{Unscaled (a) and scaled (b) fluctuations after a double quench.} \textbf{a,} We plot the unscaled fluctuations along ${y}$ direction at the predicted critical points for system sizes of ${N=10}$--${50}$ ions. 
The second quench is applied when the fluctuations following the first quench reach their maxima and the time ${t_2}$ is counted after the second quench. \textbf{b}, We apply the same scaling collapse technique as for the single quench to find the best scaling exponents (${\alpha_{2},\zeta_{2}}$) for the experimental data. See Supplementary Fig. 4 for numerical simulations of the data. We observe that the critical fluctuations do not monotonically grow for increasing system sizes, as would be expected from the scaling relations. This effect can be attributed to the imperfect switching time between the first and second quench; a nearly perfect collapse can be reproduced numerically using the precise switch times (SI Sec.~VII.D). We also report exponents found by a power-law fit of the maximum fluctuations which agree more closely with the analytical prediction (\textbf{Inset b.}). While determining the experimental critical exponents, we have excluded 50 ion data (gray) [see main text for details]. The error bars of the experimental data are statistical fluctuation around the mean value. 
} \label{fig4}
\end{figure*}

Applying the equipartition theorem, ${\lim_{t\to\infty} \Omega^2 \langle x^2\rangle_t \sim T_{\rm eff}}$ at long times, we find that the gapless mode is described by a finite effective temperature (SI Secs.~IV.A \& V.A). In fact, identifying ${\langle x^2\rangle \sim N^\alpha}$ and ${\Omega\sim N^{-\zeta}}$, the equipartition theorem reveals that the effective temperature obeys the scaling relation ${T_{\rm eff}\sim N^{\alpha - 2\zeta}}$. Now, with ${\alpha\to \alpha_1=1/2}$ and ${\zeta\to \zeta_1=1/4}$, the effective temperature becomes a constant independent of system size, consistent with thermal equilibrium behavior. Finally, we note that while the description of the soft mode in terms of the spin operators takes a more complicated form due to the experimental interaction matrix, our scaling analysis remains valid (see SI Secs.~V \& VI).

To break away from the effective thermalization, we now consider the scenario in \cref{fig1}a.~(bottom panel) where the initial state is gapless (i.e., critical) itself. 
However, realizing such a scheme in experiment can be challenging as it requires high-fidelity adiabatic preparation of the non-trivial critical state prior to the quench. In this work, we instead modify this scheme by applying a sequence of critical quenches. While the resulting critical behavior from multiple quenches is not quantum in nature, due to the over-heating of the soft mode, the prethermalization of our quantum system is crucial to evade thermalization and hence identify genuinely non-equilibrium behavior~\cite{Mori2019}.

Experimentally, we first perform a single quench (${\gamma^x=1,\gamma^y=0}$) to a critical point and evolve until the fluctuations reach their first maxima. Then we switch the interaction from the ${x}$ to the ${y}$-direction (Fig.~\ref{fig1}c) to apply a second quench i.e. we make ${\gamma^x=0,\gamma^y=1}$ (Methods). The intermediate evolution after the single quench brings the system to a critical state when the second quench is applied. Upon the second quench, the dominant fluctuations form along the ${y}$-direction and in Fig.~\ref{fig4}a, we show the unscaled fluctuations (${\mean{C_y^2}/N}$) for system sizes up to 50 ions. These fluctuations also obey the scaling relation in Eq.~\eqref{scalefunction}, but with the replacement ${C_x^2 \to C_y^2}$, ${t_1 \to t_2}$ (time after the latest quench) and with a new set of exponents ${\alpha_2}$ and ${\zeta_2}$ identifying a new universal critical behavior. We find the optimal collapse 
for ${\alpha_2=0.63\:(33)}$ and ${\zeta_2=0.10\:(17)}$ (Fig.~\ref{fig4}b). We verify that exact numerical simulation results in very similar critical exponents (Supplementary Fig.~4). Notably, these exponents are also in good agreement with the analytical values ${\alpha_2=3/4}$ and ${\zeta_2=1/8}$ obtained for the infinite-range LMG model (see SI Sec.~IV.D), underscoring the universal character of the observed critical behavior. Finally, we remark that the effective temperature now scales as ${T_{\rm eff} \sim N^{\alpha_2 -2\zeta_2}}$, which shows a nontrivial scaling with system size, underscoring a dramatic departure from equilibrium critical behavior. Such non-equilibrium criticality generally emerges upon a quench from a gapless state, and can be attributed to the over-heating of the soft mode \cite{Cazalilla_2016,Chiocchetta_2017,Paraj2020,paul2022hidden} (see SI Sec.~IV).

Experimental decoherence causes the observed fluctuations to be damped for both single and double quenches. We see that the unscaled 50 ion fluctuations after the double quench are significantly damped (Fig.~\ref{fig3}a). The major sources of decoherence, which scale with the system size, remain within acceptable thresholds for system sizes ${N<50}$, but these errors start to dominate for ${N\geq50}$ (Methods). This effect is more adverse for the double-quench sequence than the single-quench, since the former involves longer evolution under two quenches. For completeness, we have included all the 50 ion data in Fig.~\ref{fig4}a,c., but excluded it in determining the best collapse exponent. Fitting the maximum amplitudes of the fluctuations to ${N^{\alpha_2}}$ yields exponent ${\alpha_2= 0.69\:(9)}$, with tighter error bounds  (Inset Fig.~\ref{fig4}c). Errors in identifying the peak fluctuation result in erroneous switch time between the two quenches, contributing to imperfect exponents. This effect can be reproduced in the simulation with exact experimental parameters, and correction for such errors in further simulations results in exponents that are well in agreement with the analytically predicted non-equilibrium values (SI Sec.~VII.D).

\section*{Discussion}
In this work, we have demonstrated the ability to identify, both numerically and experimentally, the dynamical critical point of a disorder-to-order phase transition in a 1D transverse-field Ising model. We have observed the non-equilibrium critical behavior upon single and double quenches with up to 50 ions and extracted the universal scaling exponents, which agrees with the predictions of numerical simulation of up to 25 spins. A high level of experimental control was achieved by realizing self-similar interaction matrices across different system sizes, which is essential for experimentally identifying and characterizing the critical points of a phase transition.
While the decay of experimental spin-spin interactions deviates from the exact power-law models with $p<1$ (see SI Sec.~VII.B), the resulting universal critical behavior follows the latter models closely, a feature that is also reflected in the spin-spin correlation function (see SI Sec.~VII.C). Furthermore, we theoretically predict that the observed double-quench critical scaling is only the first in an infinite hierarchy of universal critical behaviors that emerge in a sequence of multiple quenches (SI Sec.~IV.D), an exciting direction to investigate in the future. Future work could also explore extensions of our spin-based trapped-ion simulator to non-equilibrium dynamics that include other degrees of freedom, such as phonons\cite{An2015}.

\section*{Methods}
\textbf{State preparation and readout:}
The quantum simulator used in this experiment is based on ${^{171}}$Yb${^+}$ ions trapped in all three directions in a 3-layer Paul trap \cite{Deslauriers2004} with transverse center of mass (COM) mode frequencies ranging from ${\nu_{COM}}$= ($4.64$ to $4.73$) MHz and axial COM mode frequencies ranging from ${\nu_x}$=(${0.23}$ to ${0.53}$) MHz depending on system size ({$N=10-50$}), with axial frequency being lowered to accommodate more ions in a linear chain. Before each experimental cycle, the ions are Doppler cooled in all three directions by a $369.5$ nm laser beam, ${10}$ MHz red-detuned from the ${^{2}S_{1/2}}$ to ${^{2}P_{1/2}}$ transition. We use the same laser to optically pump all the ions to initialize them in the low-energy hyperfine qubit state, ${\ket{\downarrow_z}\equiv\:^{2}S_{1/2}\ket{F=0,m_F=0}}$ with ${>99\,\%}$ fidelity. In addition to Doppler cooling, we apply the resolved sideband cooling method to bring the ions to their motional ground state with ${>90\,\%}$ fidelity. After the Hamiltonian evolution, we apply global ${\pi/2}$ rotations using composite BB1 pulses to project the spin along the ${x}$ or ${y}$ direction of the Bloch sphere from the ${z}$ direction. We then measure the magnetization of each spin using a state-dependent fluorescence by applying a beam resonant with the ${^{2}S_{1/2}\ket{F=1}\Longleftrightarrow\:^{2}F_{1/2}\ket{F=0}}$ transition. The ions scatter photons if they are projected in the ${\ket{\uparrow_z}}$ state, and appear bright, while in ${\ket{\downarrow_z}}$ state, the number of scattered photons are negligible and the ions appear dark. A finite-conjugate NA = 0.4 objective lens system (total magnification of ${70\times}$) collects scattered ${369.5}$ nm photons and images them onto an EMCCD camera, which allows us to perform site-resolved magnetization and correlation measurements with average fidelity of ${97\,\%}$. No state preparation and measurement (SPAM) correction has been applied to data presented in this work. More details of this experimental apparatus can be found in our previous works~\cite{Zhang2017DTC}~\cite{kihwan2009,Monroe2021}.

\textbf{Generating XX and YY type Ising interaction: }
The global spin-spin interaction in the trapped ion system in consideration is generated by applying a spin-dependent force via non-copropagating 355 nm pulsed laser beams that uniformly illuminate the ion chain. The pair of beams have a relative wavevector difference along the transverse motional direction of the ions. These beams are controlled by acousto-optic-modulators which impart beatnote frequencies at ${\nu_{COM}\pm \mu}$, and phases (${\phi_b}$, ${\phi_r}$), respectively, where ${\mu}$ is the symmetric detuning from the COM mode (${\approx}$ 56 kHz). These two tones respectively drive the blue (BSB) and red (RSB) sideband transitions, which, following the M\o lmer-S\o rensen (MS) protocol~\cite{MS1999}, generates an effective Hamiltonian

\begin{align}
    H=\sum_{i=1}^{N}\sum_{m=1}^{N}\frac{\eta_{i,m}\Omega_i}{2}&[a_me^{\iota\delta_mt}e^{i\phi_M}+\nonumber\\
    &a_m^\dagger e^{-\iota\delta_mt}e^{-i\phi_M}]\sigma^{\phi_s}_i,
\end{align}
where ${\eta_{i,m}}$ is the Lamb-Dicke parameter for ion ${i}$ and mode ${m}$, ${\Omega_i}$ is the Rabi frequency at ion ${i}$, ${a_m^\dagger, a_m}$ are the creation and annihilation operators of motional quanta for ${m}$th motional mode, ${\delta_m=\mu-\nu_m}$ is the MS detuning from the ${m}$th motional mode frequency ${\nu_m}$. ${\sigma^{\phi_s}_i=\cos \phi_s\sigma_i^x+\sin\phi_s\sigma_i^y}$, where the spin-phase is ${\phi_s=\frac{\phi_b+\phi_r+\pi}{2}}$ and the motional-phase is ${\phi_M=\frac{\phi_b-\phi_r}{2}}$ for the phase-sensitive realization of the MS scheme~\cite{Volkan2014}. The unitary time evolution operator under this Hamiltonian (${U(t)\sim e^{-\iota Ht}}$) can be found by taking the Magnus expansion, which after appropriate approximation leads to an effective Hamiltonian~\cite{Monroe2021}

\begin{equation}
    H=\sum_{i,j}J_{ij}\sigma^{\phi_s}_i\sigma^{\phi_s}_j.
    \label{Heff}
\end{equation}
In the far detuned limit (${\delta_m\gg\eta\Omega}$), where the virtual couplings to the phonon modes are sufficiently suppressed, the analytical form of the Ising coupling between ions ${i}$ and ${j}$ is given by~\cite{Monroe2021}

\begin{equation}\label{Jij_mat}
    J_{ij}=\Omega^2\nu_R\sum_m\frac{b_{im}b_{jm}}{\mu^2-\nu_m^2}\approx\frac{J}{\lvert i-j\rvert^p},
\end{equation}
where ${\nu_R=h\delta k^2/(8\pi ^2M)}$ is the recoil frequency, and ${b_{im}}$ is the eigenvector matrix element of the ${i}$-th ion’s participation in the ${m}$-th motional mode (${\sum_i\lvert b_{im}\rvert ^2=\sum_m\lvert b_{im}\rvert ^2=1}$), ${M}$ is the mass of the single ion. ${J}$ is the effective interaction strength obtained by a power-law fit of the interaction matrix elements, and ${J/2\pi}$ ranges within (0.25 to 0.4) kHz in the experiment for different system sizes.
If we set ${\phi_r=0,\phi_b=\pi}$, then ${\phi_M=\pi/2}$ and ${\phi_s=\pi}$, which makes the Hamiltonian of Eq.~\eqref{Heff} an effective ${\sigma^x\sigma^x}$ interaction. We can change this phase by changing the input waveform to the acousto-optic-modulator. Similarly, we set ${\phi_r=0,\phi_b=0}$ to obtain an effective ${\sigma^y\sigma^y}$ interaction. In the double quench experiment, we switch these waveform phases to switch between interactions along different Bloch sphere directions. 

We further apply a common offset of ${2B^z}$ to the frequencies of BSB and RSB tones which in the rotating frame of the qubit, results in an effective transverse field term ${B^z\sum_{i}^{N}\sigma^z_i}$ in the Hamiltonian of Eq.~\eqref{Heff}~\cite{Monroe2021}. The magnetic field strength ${B^z}$ is chosen such that ${B^z\ll\delta_m}$ for the rotating frame approximation to be valid.

The approximate power law exponent can be theoretically tuned within the range ${0<p<3}$. However, in this experiment, we kept the exponent ${\approx 0.89}$ for all the system sizes by tuning the axial trap frequency (${\nu_x}$) and motional detuning (${\mu}$). We also note that the experimental interaction matrix deviates from a pure power-law decay to an exponential decay at large distances, especially for large system sizes (SI Sec.~VIII).  In principle,  one can tune this exponent by changing only the detuning (${\mu}$) (see Eq.~\eqref{Jij_mat}), but changing the axial trap frequency (${\nu_x}$) for different system sizes results in more self-consistent scaling of the exact spin-spin coupling matrix~\cite{Pagano2020}. We remark that maintaining a self-similar interaction matrix can also be achieved with individual optical control over all the ions~\cite{Feng2023}, but it is beyond the scope of this experimental setup, which features global optical control.

\textbf{Experimental error sources:}
One of the main challenges in scaling up the system size is to maintain the fidelity of the quantum simulation experiments. Among various sources of decoherence in the trapped-ion platform, such as stray magnetic and electric fields, mode frequency drifts, off-resonant motional excitation, spontaneous emission, and additional spin-motion coupling that causes the evolution to depart from ideal simulation\cite{Monroe2021}. One such important source, which becomes significant in the large system size limit, is the off-resonant excitation of the motional modes causing residual spin-motion entanglement~\cite{Pagano2020,tan2021domain}. In order to trap longer linear chains while maintaining the same interaction profile, we need to operate at a lower axial confinement which can become as low as ${\sim 200}$ Hz for ${N=40-50}$. At such low axial confinement, the trapped ions are more susceptible to electric field noise and background collisions~\cite{Marko2022}. The conventional laser cooling methods start to become inefficient in cooling the ions to their motional ground states and as a result, errors due to phonon evolution get introduced in the Hamiltonian evolution. To the lowest order, such an error can be modelled as an effective bit flip error during measurement~\cite{Pagano2020}. Additional cooling methods such as EIT (electromagnetically induced transparency) cooling~\cite{Feng2020} and sympathetic cooling~\cite{Marko2022} would be useful in mitigating the effects of such errors. 

Another source of bit-flip error is imperfect detection. Off-resonant pumping from the detection beam limits our detection fidelity to about ${98\,\%}$. When a large number of ions are trapped in a linear chain, ions near the center of the chain are closer together than the ones at the edges. A random bit-flip error can be introduced due to leakage of light from neighbouring ions, which might cause a dark ion to appear bright and vice versa. More details about various noise sources in this apparatus can be found in previous works~\cite{kihwan2009,Pagano2020,Morong2022}.

\textbf{Modeling decoherence:} To account for the most relevant source of experimental dissipation, we consider an effective model that incorporates bit-flip errors in the dynamics. In this model, the density matrix evolves in time under the Lindblad master equation as
\begin{equation}
    \frac{d\rho}{dt} = -i[H,\rho] +  \kappa \sum_{i=1}^N (\sigma_i^x \rho \sigma_i^x -\rho) ,
    \label{eq:linblad}
\end{equation}
where the single bit-flip rate $\kappa$ is chosen such that the numerically obtained net correlators $\langle C_x^2 \rangle$ best fit the experimental data. For simplicity, we have chosen the bit-flip rate to be uniform across the chain. In the simulations, we numerically calculate the dynamics under \cref{eq:linblad} using experimental Hamiltonian parameters and the best-fit value of $\kappa/\mathcal{J}$ for a given $B^z/\mathcal{J}$. For ${N=10}$, we find the best fit at $B^z/\mathcal{J}=\left[0.21, 0.50, 0.87, 1.00, 1.50, 1.93\right]$ to be $\kappa/\mathcal{J} = \left[0.057, 0.036, 0.089, 0.085, 0.067, 0.045\right]$, respectively. For ${N=15}$, and $B^z/\mathcal{J}=1$, we have $\kappa/\mathcal{J}=0.024$. In Supplementary Fig.~1, we present the experimentally observed dynamics (represented by dots) of the net correlator after a single quench along with their numerical simulation (represented by solid lines) with 10 and 15 ions. In the subfigure a (${10}$ ions), different colors indicate different values of the quench parameter $B^z/\mathcal{J}$. In subfigure b, we show the quench to $B^z/\mathcal{J}=1$ for ${N=10}$ and ${15}$. As shown in Supplementary Fig. 1, the numerical simulations show good agreement with the experimental data, capturing the damping rate as well as the frequency of the underdamped oscillations. The numerical results for ${10}$ ions are obtained under the exact time evolution of the Lindblad master equation using QuTiP's implementation\cite{qutip}. Simulations for systems with ${15}$ ions are obtained using the Monte Carlo wave function method\cite{qmc}. This method can, in principle, be used to simulate larger system sizes. However, optimizing the decoherence rate in order to fit the experimental data requires a large number of these simulations, thus limiting our ability to simulate larger system sizes.

\textbf{Jackknife error estimation:} In the experiments reported in this work, we repeat the experiment and measurement sequence ${400}$ times to reduce the quantum projection noise. To estimate the standard errors of the two-body correlators, we have implemented a Jackknife resampling technique~\cite{MillerJK}. In this method, we construct a distribution of net correlators by randomly sampling 399 experimental runs, each time leaving out only one run. We then calculate the variance of the distribution which corresponds to the standard error of the net correlator.

\section*{Data Availability}
All key experimental data supporting the findings of this study are provided in the main text and Methods. The raw datasets are available from the corresponding authors upon request for purposes of academic replication and review. Access is limited due to the large size and specialized formatting of the data, which requires additional processing to ensure correct interpretation.

\section*{Code availability}
The code used for analyses is available from the corresponding author upon request.

\section*{Acknowledgement}
We acknowledge insightful discussions with W.L. Tan and L. Feng. This material is based upon work supported by the U.S. Department of Energy (DOE), Office of Science, National Quantum Information Science Research Centers, Quantum Systems Accelerator. Additional support is acknowledged from the NSF STAQ Program (PHY-1818914), the NSF QLCI (award No.~OMA-2120757), and the AFOSR MURI on Dissipation Engineering in Open Quantum Systems (FA9550-19-1-0399). 
M.A., D.P. and M.M are supported by the AFOSR (FA9550-20-1-0073) and the NSF CAREER Award through DMR-2142866.
P.T. acknowledges funding from DOE ASCR (ERKJ347).
P.C. was supported by the Department of Defense (DoD) through the National Defense Science and Engineering Graduate (NDSEG) Fellowship Program. 
G.P. acknowledges that this material is based upon work supported by the U.S Department of Energy, Office of Science, Office of Nuclear Physics under the Early Career Award No. DE-SC0023806.

\section*{Author contributions}
A.D., W.M., K.C., and C.M. contributed to the experimental design, construction, data collection and analysis. G.P. provided experimental support. P.C., M.A., D.P., P.T., A.V.G., and M.M. carried out the theoretical analysis. All authors contributed to the discussion of the results and the manuscript.

\section*{Competing interests}
All the authors declare no competing interests.

\end{document}


\title{ Supplementary Information}

\author{Arinjoy De$^{*,\dagger}$~\orcidlink{0000-0001-9184-8434}}
\affiliation{\JQI}
\author{Patrick Cook$^*$}
\affiliation{\MSU}
\affiliation{\FRIB}
\author{Mostafa Ali}
\affiliation{\MSU}
\author{Kate Collins}
\affiliation{\JQI}
\author{William Morong $^\ddagger$}
\affiliation{\AWS}
\author{Daniel Paz}
\affiliation{\MSU}
\author{Paraj Titum}
\affiliation{\JQI}
\affiliation{\JHU}
\author{Guido Pagano}
\affiliation{\RICE}
\author{Alexey V. Gorshkov}
\affiliation{\JQI}
\author{Mohammad Maghrebi}
\affiliation{\MSU}
\author{Christopher Monroe~\orcidlink{0000-0003-0551-3713}}
\affiliation{\JQI}
\affiliation{\DUKE}

\maketitle
\thispagestyle{firstpage}

\tableofcontents
\vspace{1cm}

This document provides additional data figures and numerical simulations relevant to the main text (\cref{sec:additional data}). \Cref{sec:into} introduces the theoretical model and the following  sections detail the theoretical analysis (\cref{sec:meanfield-analysis,sec:LMG,sec:long-range,sec:experiment_long-range}), elaborates on and contrasts against the experimental results (\cref{sec:expt_vs_theory}), 
and finally provides the numerical approach used to determine the critical exponents and their errors (\cref{sec:optimal_collapse}). 

\section{Additional data figures and numerical simulations:}\label{sec:additional data}

\begin{figure}[H]
   \centering
   \includegraphics[width=1\linewidth]{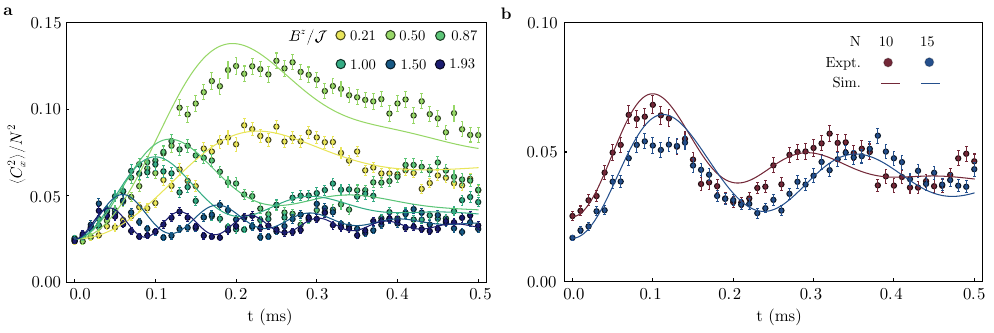}
   \caption{\textbf{Net correlator dynamics including decoherence effects:} Here we compare the evolution of the experimental (dots) net correlator (${\mean{C_x^2}/N^2}$) after a single quench with numerical simulation. Due to decoherence and imperfect detection fidelity, the fluctuations in the experimental data are damped compared to the ideal simulation. This effect is captured in the Lindbladian master equation (Eq.~7 in Methods) by incorporating a noisy bit-flip term along the $x$ direction in the numerical simulations of the Hamiltonian (solid lines), which agrees well with the experimental data (see Methods). The error bars represent statistical standard error around the mean. \textbf{a,} 10 ions for multiple values of $B^z/\mathcal{J}$: different colors represent the evolution at different values of ${B^z/\mathcal{J}}$. The net correlator increases in amplitude and evolves more slowly as ${B^z}$ is swept from larger to smaller values, except near ${B^z/\mathcal{J}\rightarrow 0}$ where there are no correlations. \textbf{b,} Dynamics of 10- and 15-ion chains after a single quench to $B^z/\mathcal{J}=1$.}
   \label{fig5}
\end{figure}

\begin{figure}[H]
    \centering
    \includegraphics[width=0.6\linewidth]{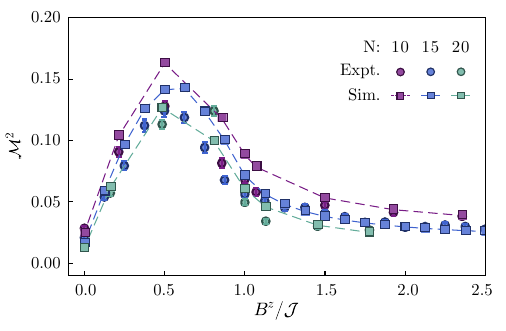}
    \caption{\textbf{Scaled maximum net correlator comparison :} Here we show the comparison of the experimental and numerical net maximum correlator ${{\cal M}^2=\text{Max}\left [\mean{C^2_x }/N^2\right ]}$ as a function of scaled magnetic field $B^z/\mathcal{J}$. The filled circles represent the experimental data points while the squares are obtained by exact numerical simulation of the dynamics using the long-range spin-spin couplings of the experimental system, without considering the decoherence effects. Dashed lines connecting the simulation data points are included to guide the eye.}
    \label{fig6}
\end{figure}

\begin{figure}[H]
\centering
\includegraphics{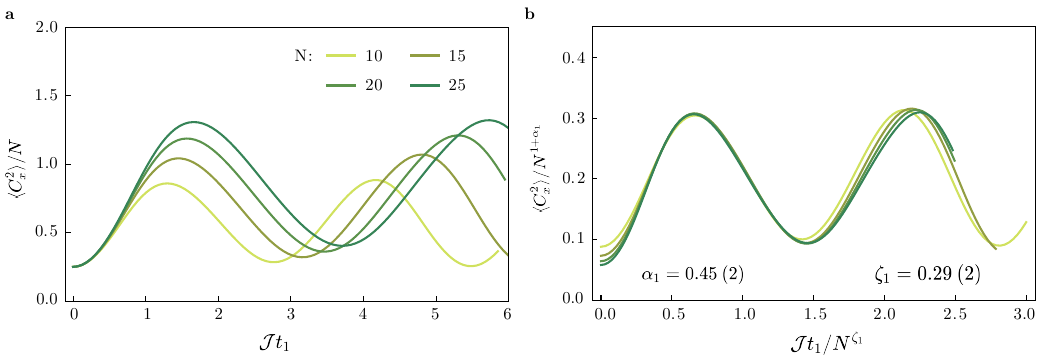}
\caption{\textbf{Numerically calculated unscaled (a) and scaled (b) fluctuations after a single quench.}  \textbf{a,} Unscaled numerical simulation of critical fluctuations, obtained from decoherence-free Hamiltonian evolution (via exact numerics) with experimental parameters for up to ${N=25}$ ions. \textbf{b,} We scale the numerical fluctuations using the optimal critical exponents (${\alpha_{1},\zeta_{1}}$), which are obtained by optimizing the numerical data independent of the experimental results. 
We observe remarkable similarity between the exponents found in the experiment (Fig. 3 in main text) and simulation, highlighting the universality of the exponents despite experimental imperfections as well as finite-size effects.} \label{fig7}
\end{figure}

\begin{figure}[H]
\centering
\includegraphics{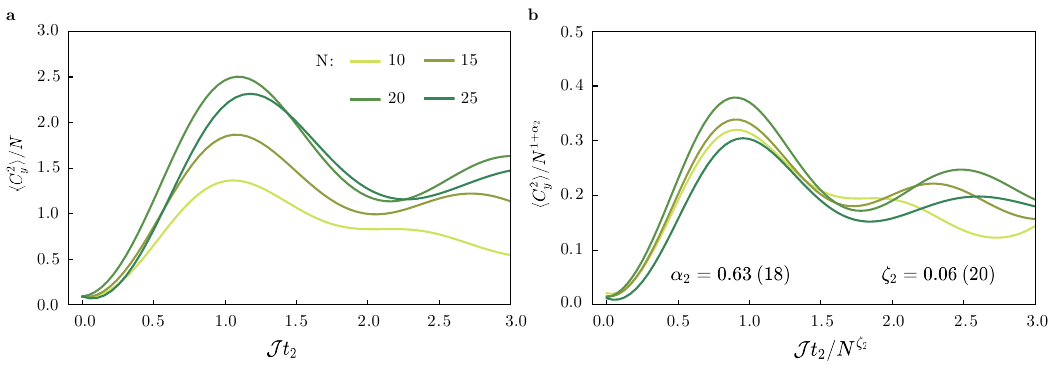}
\caption{\textbf{Numerically calculated unscaled (a) and scaled (b) fluctuations after a double quench.} \textbf{a,} Double-quench dynamics simulated using exact numerics with the experimental spin-spin coupling parameters for system sizes ranging from ${N=10}$ to ${N=25}$. \textbf{b,} We apply the same scaling collapse technique to the numerical data as for the single quench and find the best scaling exponents (${\alpha_{2},\zeta_{2}}$). We find similar exponents from the experimental data in Fig. 4 of main text.
} \label{fig8}
\end{figure}

Here we present additional data and numerical simulations that support our main findings. Supplementary~\cref{fig5} compares experimentally observed dynamics after a single quench to numerical simulation, incorporating decoherenece effects. Supplementary~\cref{fig6} shows the net maximum correlator ${{\cal M}^2=\text{Max}\left [\mean{C^2_x }/N^2\right ]}$ as a function of scaled magnetic field $B^z/\mathcal{J}$ compared against numerical predictions. Supplementary~\cref{fig7} illustrates the fluctuation dynamics after a single quench and its critical scaling up to the largest system sizes accessible by exact numerical methods. Supplementary~\cref{fig8} represents analogous results after a double quench. 

\newpage
\section{Theoretical Models}\label{sec:into}

In the following parts of the Supplemental Information, we consider theoretical models with a generalized long-range interaction: 

\begin{equation}\label{eq:H_generalized}
    H
    = -\frac{1}{2\mathcal{J}}\sum_{i\neq j}^N \Jij\left(\gx \sigma_i^x \sigma_j^x + \gy \sigma_i^y \sigma_j^y\right) + B \sum_{i}^N \sigma_i^z,
\end{equation}
where $\sigma_i^\alpha$ with $\alpha=x,y,z$ are the usual Pauli operators for spin $i$, and

\begin{equation}
    \mathcal{J} = \frac{1}{N-1}\sum_{i\neq j}^N \Jij,  
\end{equation}
is a normalization constant known as the Kac factor (or, Kac normalization).  This normalization factor is particularly useful for long-range coupling as it renders the Hamiltonian extensive.  

\begin{enumerate}
    \item In \cref{sec:meanfield-analysis}, we provide a mean-field analysis to determine the dynamical phase diagram. We then consider three versions of the model described by \cref{eq:H_generalized} ranging from the most idealized theoretical model to one that is most experimentally relevant.
    \item In \cref{sec:LMG}, we consider the collective Lipkin-Meshkov-Glick (LMG) model where $J_{ij}=1$ and the interaction is infinite ranged. We provide a detailed analysis of the ground-state, the dynamical phase diagram as well as the critical properties. While an idealized model, it explains the main features of the experimental results. 
    \item In \cref{sec:long-range}, we add another layer of complexity by considering a power-law interaction $J_{ij} \sim 1/|i-j|^p$ with open and periodic conditions. We show that, for $0<p<1$, this model exhibits the same dynamical phase diagram and critical properties. 
    \item Finally, in \cref{sec:experiment_long-range}, we theoretically analyze the long-range spin model with the experimental values of $J_{ij}$. The latter values fall off exponentially at large distances (see \cref{sec:Jij comparison}), hence departing from the collective or the truly long-range variant of the previous models. Nonetheless, we show that the critical properties of the experimental model are consistent with those of the LMG model and its long-range variants. 
\end{enumerate}

Before continuing, we note that the experimental model is antiferromagnetic while the theoretical models considered here are ferromagnetic; however, we use a simple trick to relate the two models \cite{Islam2013,Jurcevic2017,tan2021domain}. Given that the Hamiltonian is purely real (using the standard representation of Pauli matrices) and the fact that the correlation function $\langle \sigma^x_i \sigma^x_j\rangle$ is real, we have 

\begin{equation}\label{eq:sigma_x_correlator}
    \langle \sigma^x_i \sigma^x_j\rangle_{H, \psi_0, t} =      (\langle \sigma^x_i \sigma^x_j\rangle_{H, \psi_0, t})^*=\langle \sigma^x_i \sigma^x_j\rangle_{-H, \psi_0, t},
\end{equation}
where we have made explicit the dependence on the Hamiltonian $H$, time $t$ and the initial state $\psi_0$.
We have furthermore assumed that the initial state is also real; for example, a fully polarized state along the negative $z$ direction is real. By flipping the overall sign of the Hamiltonian, the antiferromagnetic coupling is then mapped to a ferromagnetic one. 

\section{Dynamical Phase Diagram from Mean-Field Analysis}\label{sec:meanfield-analysis}
In this section, we obtain the dynamical phase diagram using a mean-field analysis for the generalized model in \cref{eq:H_generalized}. Starting from the initial state $\ket{\downarrow \downarrow \cdots}$, we suddenly turn on an Ising-type Hamiltonian by setting $\gamma^x=1$ and $\gamma^y=0$. In a mean-field analysis, the Hamiltonian becomes 

\begin{equation}
    H= - m  \sigma^x +B\sigma^z =\sqrt{m^2+B^2}\,\, {\mathbf n} \cdot {\boldsymbol\sigma},
\end{equation}
where $m\equiv \langle \sigma^x\rangle$. We have dropped the site index, and defined the unit vector $\mathbf n = (n_x, 0,n_z)$ parallel to $(-m, 0, B)$. At long times after the sudden quench, the initial state  dephases in the new basis defined by the operator ${\mathbf n} \cdot {\boldsymbol\sigma}$, and the time-averaged density matrix becomes

\begin{equation}
    \lim_{t\to \infty}|\psi\rangle_t\langle\psi| = \sin^2 \frac{\theta}{2}|{\mathbf n},+\rangle \langle{\mathbf n},+| + \cos^2\frac{\theta}{2} |{\mathbf n},-\rangle \langle{\mathbf n},-|,
\end{equation}
where $\mathbf n \cdot \mathbf z=\cos\theta$. Using the identity $m=\langle \sigma^x\rangle$, we then find the self-consistent equation

\begin{equation}
     m = -\cos \theta \sin \theta = \frac{ m}{\sqrt{m^2+B^2}}\frac{B}{\sqrt{m^2+B^2}},
\end{equation}
which in turn yields the time-averaged order parameter as

\begin{equation}
    \label{eq:mean-field}
     m= \sqrt{B\left(1-B\right)}.
\end{equation}
Furthermore, this equation predicts a dynamical phase transition at $B_c=1$. 
This mean-field analysis is exact for the infinite-range LMG model. Furthermore, with the aid of spin-wave analysis, we show that it is also exact for the long-range model with $0<p<1$ (see \cref{sec:long-range}), and additionally provides an excellent estimate for the phase transition corresponding to the experimental $J_{ij}$ (see \cref{sec:experiment_long-range}). 

\section{LMG model}\label{sec:LMG}
The LMG model is characterized by the infinite-range interaction with $\Jij\equiv 1$. This model can be reformulated in terms of total spin operators, with the Hamiltonian (up to a constant)

\begin{equation}\label{eq:LMG}
    H_{LMG}  =  
    - \frac{2}{N}\left(\gx S_x^2 + \gy S_y^2\right) + 2B S_z,
\end{equation}
where $S_\alpha= \frac{1}{2}\sum_i\sigma^\alpha_i$. At zero temperature, this model exhibits a quantum phase transition at $B_{gs, c} = 1$ \cite{suzuki2012quantum}.
Interestingly, the ground state critical point coincides exactly with the dynamical critical point obtained in the previous section. However, there are important distinctions: the order parameter's dependence on $B$ is different. In fact, the ground state is perfectly ordered at $B=0$ while the corresponding order parameter long after the sudden quench vanishes when $B=0$. More importantly, the critical behaviour at the respective phase transitions is different: while the ground state exhibits a quantum phase transition at $B=1$, the dynamical critical point shows distinct (and, effectively thermal) critical behaviour (see \cref{sec:critical_LMG}). Furthermore, we show that for a double quench, the critical behaviour is neither quantum nor thermal, and is genuinely non-equilibrium (see \cref{sec:double_quench}).

\subsection*{Beyond Mean Field}
Next, we consider fluctuations at the phase transition. While the mean field solution in \cref{eq:mean-field} exactly describes the ordered phase (where $B<B_c =1$) in the thermodynamic limit, fluctuations and finite-size scaling become important at or near the critical point for a finite system. To characterize fluctuations in the disordered phase as well as critical fluctuations near the critical point, one can use the Holstein-Primakoff transformation to map the spin operators of the LMG model to bosonic creation and annihilation operators as ($S_\pm=S_x+iS_y$)

\begin{equation}
    S_-  = S_+^\dagger =\sqrt{N-a^\dagger a}a \qquad S_z = -N/2 + a^\dagger a.
\end{equation}
To the lowest order in $1/N$, we can identify

\begin{equation}\label{eq:HP}
        S_x \approx \sqrt{\frac{N}{2}}x, \qquad S_y \approx \sqrt{\frac{N}{2}}p.
\end{equation}
Assuming that $\gamma^x>\gamma^y$, the dominant Ising interaction is along the $x$ direction. In this case, the LMG Hamiltonian can be simply written as that of a harmonic oscillator \cite{Titum20},

\begin{equation}
\label{eq:HP_LMG_full}
H = \frac{1}{2m}p^2 + \frac{1}{2}m\Omega^2x^2 + \frac{1}{4N}u x^4 + \cdots,
\end{equation}
where we have defined $m^{-1} = 2\left(B - \gy\right)$, $\Omega^2 = 4\left(B - \gx\right)\left(B - \gy\right)$, and $u = 2\gx$. The dots in the above equation represent higher-order terms in $1/N$ or higher powers of $x$ and $p$ which are irrelevant at or away from the critical point in the disordered phase. 

\subsection{Gaussian Fluctuations}\label{sec:effectiveTemp}
We begin by considering \cref{eq:HP_LMG_full} at the quadratic level in the disordered phase, before taking into account the interaction term ($\propto 1/N$). We are interested in the dynamics upon the sudden quench $(\gamma^{x,y}_0, B_0)\to (\gamma^{x,y},B)$. Equivalently, we can consider a quench $(m_0,\Omega_0) \longrightarrow (m,\Omega)$ of the harmonic oscillator, where it is easy to solve for the time-averaged fluctuations:

\begin{equation} 
    \frac{1}{N}\overline{\langle S_x^2\rangle}=\frac{1}{2}\overline{\expectation{x^2}} = \frac{m_0\Omega_0}{8 m^2 \Omega^2} \left(1+\frac{m^2\Omega^2}{m_0^2\Omega_0^2}\right)\,,
\end{equation}
where the overline indicates time averaging. Notice that fluctuations diverge where $\Omega=0$, which in turn sets $(B-\gamma^x)(B-\gamma^y)=0$. This condition coincides with the mean-field prediction in \cref{eq:mean-field} when $\gamma^x=1$ and $\gamma^y=0$; this is expected as the LMG model is infinite ranged. Furthermore, we can characterize the critical fluctuations beyond the mean-field prediction. We first observe that fluctuations scale as $\overline{\langle x^2\rangle} \sim 1/\Omega^2$ near the critical point. 
Comparing against the expression for a harmonic oscillator at high temperatures 

\[
    \frac{1}{2}m\Omega^2 \expectation{x^2}\stackrel{T\gg\Omega}{\approx} \frac{1}{2}T,
\]
we can identify an \textit{effective temperature} \cite{Titum20}

\begin{equation}\label{eq:T_efff}
        T_{\mathrm{eff}} = \frac{m_0\Omega_0}{4 m}=\frac{B}{2}\sqrt{\frac{B_0-\gamma^x_0}{B_0-\gamma^y_0}}.
\end{equation}
Specifically, in a quench starting from a product state corresponding to the ground state in the absence of the Ising interaction ($\gamma^{x,y}_0=0$) to the critical point of the infinite-range Ising Hamiltonian ($\gamma^x=1$, $\gamma^y=0$, and $B=1$), we find a constant effective temperature, $T_{\rm eff} = 1/2$.  We stress that such notion of the effective temperature does not mean that the system is in equilibrium; rather, it means that low-frequency modes (or, more precisely, the soft modes) have become effectively thermal. Indeed, one finds thermal critical behaviour and scaling in this type of quench \cite{Titum20}; see also Supplementary~\cref{fig:scaling}.

\subsection{Critical Finite-Size Scaling}\label{sec:critical_LMG}
Fluctuations diverge as we approach the critical point $B\to 1$ either in the ground state or in a quench to the critical point. In a finite-size system, the fluctuations do not strictly diverge but rather scale with $N$ in a nontrivial fashion:

\begin{equation}
    \frac{1}{N}\langle S_x^2\rangle \sim N^\alpha.
\end{equation}
Notice that in the disordered phase, we have $\alpha=0$ indicating noncritical fluctuations while in the ordered phases $\alpha=1$ due to ordering $\langle S_x\rangle\sim N$. Only at the critical point, $\alpha$ becomes a nontrivial critical exponent. For the quantum phase transition in the ground state, this exponent becomes $\alpha=1/3$; see Supplementary~\cref{fig:scaling}(a). This should be contrasted with a thermal phase transition where this exponent becomes $\alpha=1/2$. In a quench to the critical point, the system evolves and approaches a stationary state that is not an equilibrium state; however, the critical exponent takes the same value $\alpha=1/2$ as thermal equilibrium; see Supplementary~\cref{fig:scaling}(b). One can view this as a consequence of the fact that the effective temperature takes a finite nonzero value (see \cref{eq:T_efff}). Here, we derive this critical scaling using a simple analysis. Our approach is based on a non-equilibrium field theory which systematically leads to a semiclassical analysis similar to the truncated Wigner approximation (see Ref.~\cite{Titum20} for more details). 
To this end, we start from the partition function

\begin{equation}
    Z= \int Dx_\pm W_0 e^{i(S[x_+]-S[x_-])}\,,
\end{equation}
where $W_0$ is the Wigner function corresponding to the initial state, and $[x_\pm(t)]$ denote forward and backward trajectories, respectively. These trajectories are weighted by the phase factors $\exp(\pm iS)$ where  $S=\frac{1}{2}\int dt[m\dot x^2-m\Omega^2 x^2 - (u/2N)x^4]$ is the corresponding action. Introducing the (Keldysh) variables $x_{c/q}=(x_+\pm x_-)/\sqrt{2}$, the total (Keldysh) action becomes 

\begin{equation}
    S_K = S[x_+] - S[x_-] =-\int dt [mx_q \ddot x_c + r x_q x_c + (u/2N) (x_c^3 x_q + x_c x_q^3)]\,,
\end{equation}
with $r=m\Omega^2$. We take the initial state as the ground state of the Hamiltonian in the absence of the Ising interaction. The Wigner function then becomes a Gaussian function of phase  space variables as

\begin{equation}
    W_0(x_0, p_0) \sim e^{-x_0^2- p_0^2}\,.
\end{equation} 
Casting the Wigner function in terms of the configuration variables, the total partition function takes the form

\begin{equation}
    Z= \int Dx_{c/q} e^{-x_{c0}^2/2l_0^2 - \dot x_{c0}^2/2v_0^2} e^{-\int dt [mx_q \ddot x_c + r x_q x_c + (u_c/2N) x_c^3 x_q + (u_q/2N)(x_c x_q^3)]}\,,
\end{equation}
where we have introduced the parameters $l_0$ and $v_0$ for later convenience; for the coherent initial state assumed above, we have $l_0=1$ and $v_0=1/m_0$. We have also introduced the coefficients $u_{c/q}$ to distinguish the two interaction vertices; 
at the microscopic level, we have 
$u_c= u_q=u$. As pointed out in Ref.~\cite{Titum20}, to probe the behaviour at long times, we can rescale time $t\to t/\lambda$ and identify how various terms scale. We find that that under this rescaling $l_0 \to0$ and $u_q \to0$. The first condition imposes the boundary condition $x_{c0}=0$ given the Gaussian distribution $\exp(-x_{c0}^2/2l_0^2)$. The latter condition simply means that the `quantum' vertex proportional to $u_q$ is less relevant than the classical vertex proportional to $u_c$, a condition that typically emerges in the semiclassical limit. Dropping the quantum vertex, we can integrate over the field $x_q$ to find a delta function that imposes the semiclassical equation of motion together with a Gaussian distribution of the initial state (also changing the coordinate back to $x$ via $x_c =\sqrt{2} x$ and restoring $u_c\to u$):

\begin{equation}\label{eq:Z_delta_fn}
    Z=\int Dx \delta(x_{0})e^{-\dot x_{0}^2/v_0^2} \delta\left(m\ddot x+rx+\frac{u}{N}x^3\right)\,.
\end{equation}
This equation admits a simple interpretation: A `particle' whose position is given by $x$ moves under a nonlinear equation of motion with the initial conditions where $x_{0}=0$ and $\dot x_{0}$ is drawn from a Gaussian distribution. This equation is simply the truncated Wigner approximation, which evolves the wavefunction with the classical equation of motion, but captures the (quantum) fluctuations of the initial state; the only difference is that, using a scaling argument, we have set $x_{0}=0$.

To characterize the critical behaviour, we can study the scaling behaviour of the above equation. We remark that, at the critical point $r=0$, the partition function becomes scale invariant under the rescaling 

\begin{equation}
    t\to t/ \lambda, \quad x \to x/\lambda, \quad N \to  N/\lambda^4\,.
\end{equation}
This immediately implies that fluctuations $\langle x^2\rangle$ as a function of $t$ and $N$ both take the scaling form

\begin{equation}
    \langle x(t)^2\rangle = N^\alpha f(t/N^\zeta)\,,
\end{equation}
with the critical exponents $\alpha=1/2$ and the $\zeta=1/4$; the latter exponent identifies a dynamical critical exponent. This is indeed consistent with the exact numerical calculation in Supplementary~\cref{fig:scaling}(b). 

\begin{figure}[H]
\centering
    \subfloat[]{\includegraphics[height=0.305\linewidth]{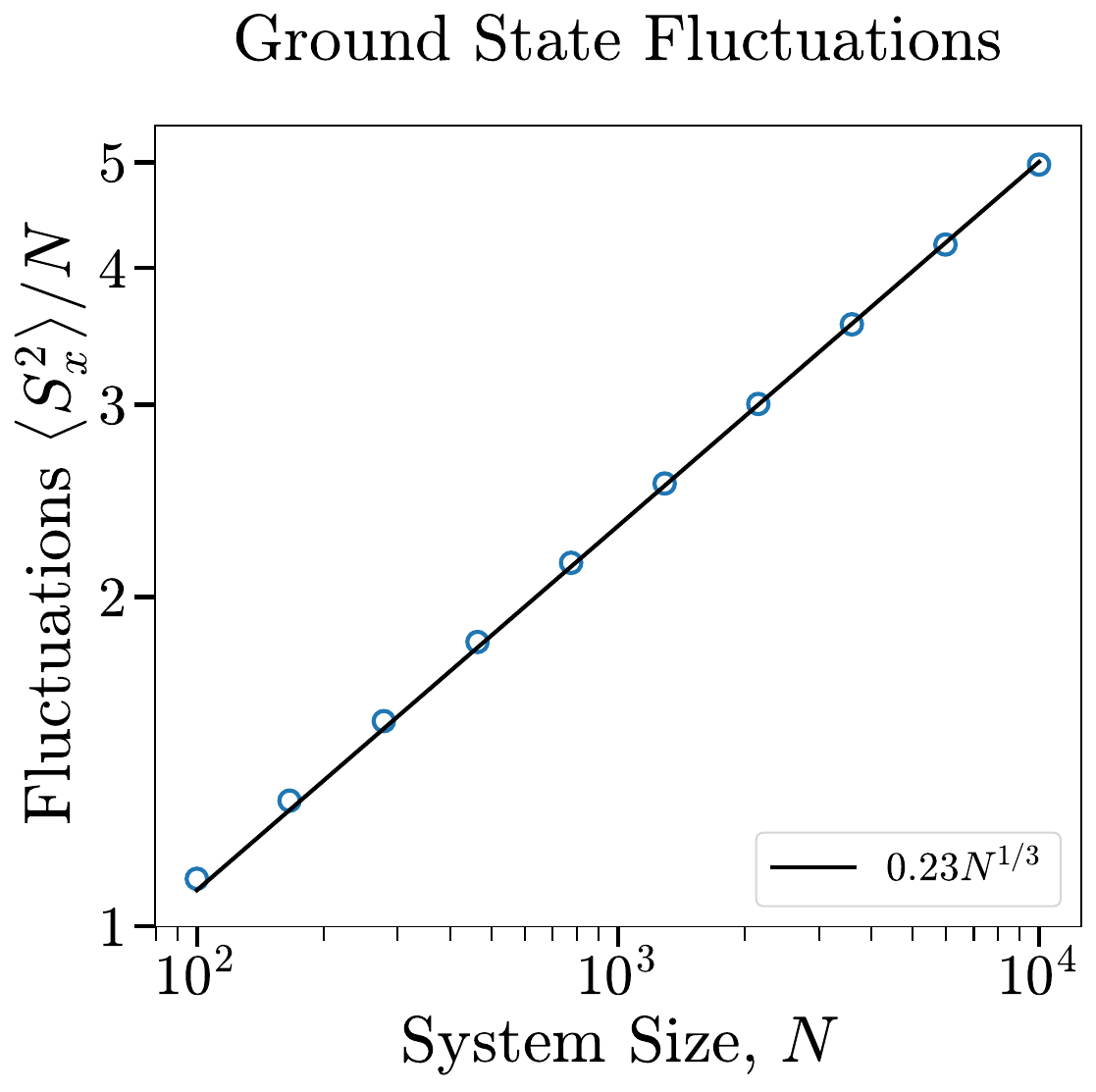}}
    \hfill
    \subfloat[]{\includegraphics[height=0.3\linewidth]{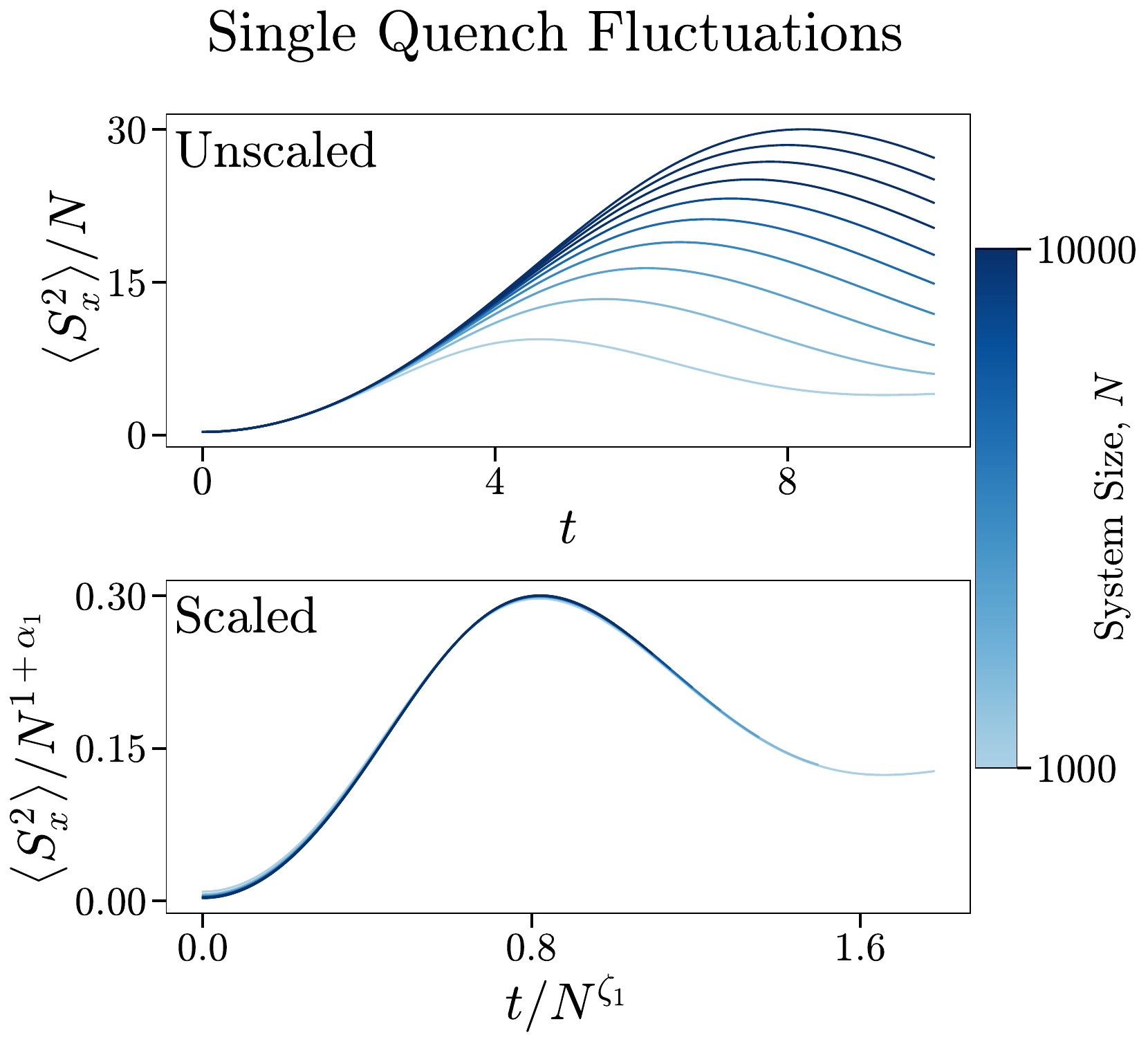}}
    \hfill
    \subfloat[]{\includegraphics[height=0.3\linewidth]{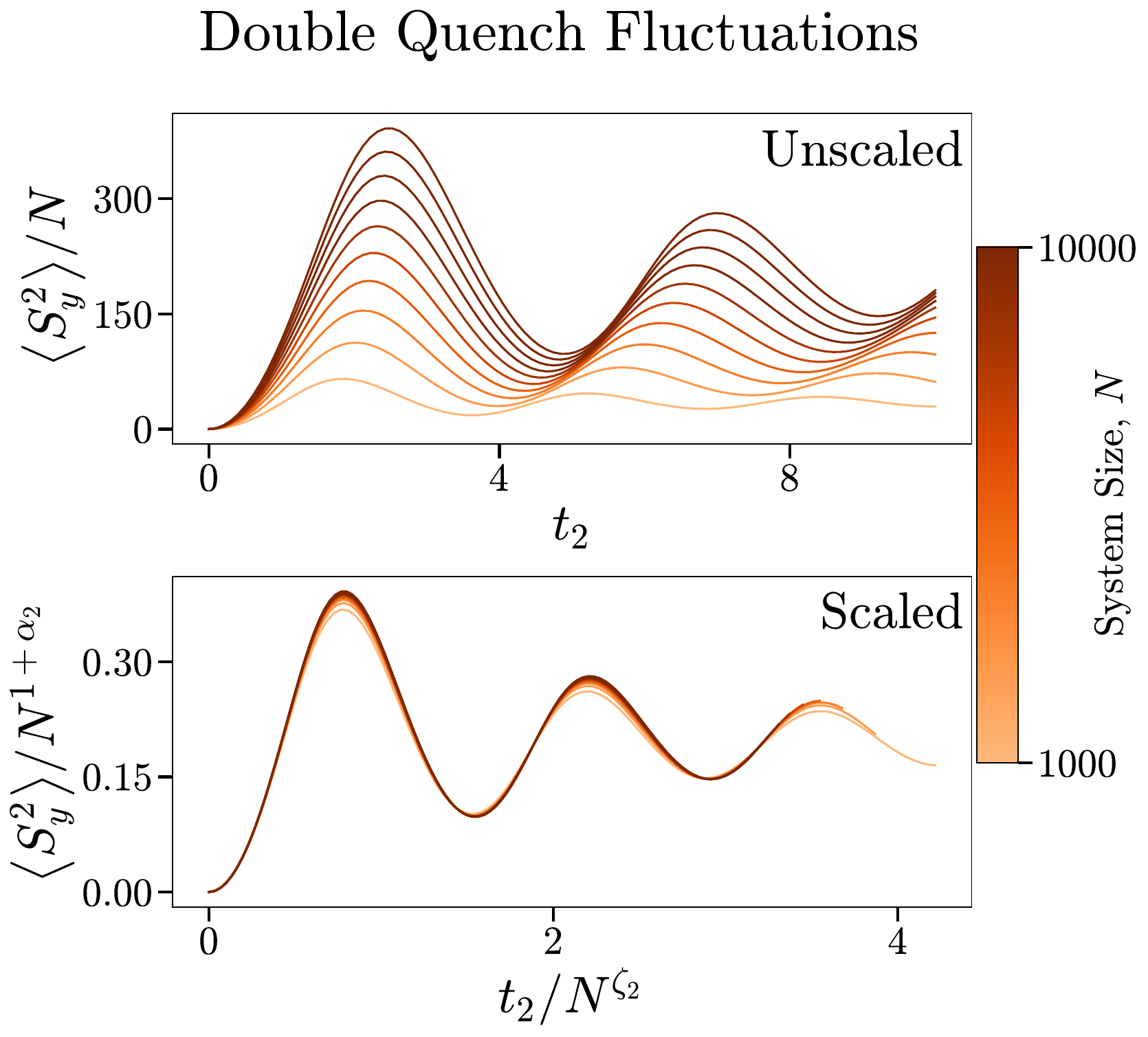}}
    \caption{Fluctuations scaling with system size at the critical point (a) in the ground state, (b) after a single quench, and (c) after a double quench. (b) A single quench from the disordered phase to the critical point ${\left(B,\gx,\gy\right) = (1,1,0)}$ gives rise the critical exponents ${\left(\alpha_1, \zeta_1\right) = \left(1/2,1/4\right)}$. (c) A second quench to the other critical point ${\left(B,\gx,\gy\right) = (1,0,1)}$ is performed when the fluctuations in panel (b) reach their peak. The coordinate $t_2$ denotes the time since the second quench. The double quench gives rise to the critical exponents ${\left(\alpha_2, \zeta_2\right) = \left(3/4,1/8\right)}$.}
    \label{fig:scaling}
\end{figure}

\subsection{Finite-Size Behavior of the Order Parameter}\label{sec:finitesize_order_param}
 In \cref{eq:mean-field}, we have presented the mean-field prediction for the time-averaged order parameter in the ordered phase, $B< 1$, and in the thermodynamic limit $N\to \infty$. In the previous section, we have further investigated the critical fluctuations for finite system sizes \textit{exactly }at the critical point $B=B_c=1$. In this section, we study finite-size corrections in the \textit{vicinity} of the critical point $B_c$. This is crucial to identify the dynamical phase transition in a finite-size system.
 To this end, we consider the equation of motion dictated by \cref{eq:Z_delta_fn}: 
 
\begin{equation}
     \ddot x + r x + u x^3 =0. 
\end{equation}
For notational convenience, we have set $m=1$ and absorbed the factor proportional to $1/N$ into the definition of the interaction parameter $u$. For the purpose of this section, it suffices that

\begin{equation}
    r\propto B-1, \qquad u \propto 1/N\,.
\end{equation}
In the quench starting from a disorded state, the initial conditions are set by $x(t=0)=0$ and $\dot x(0)=v$ where $v$ is drawn from a Gaussian distribution $\exp(-Dv^2)$; the parameter $D$ is set by the initial state, but we shall treat it as a phenomenological parameter. Solving the above equation with the initial conditions $x(t=0)=0$ and $\dot x(t=0)=v$, we find

\begin{equation}
    x_v(t)=\mathrm{sgn}(v) \, {\mathrm{sn}}\left(t \sqrt{\frac{r+\sqrt{r^2+2u v^2}}{2}}, \frac{r-\sqrt{r^2+2u v^2}}{r+\sqrt{r^2+2u v^2}}\right)\,,
\end{equation}
with ${\rm sn}(u,m)$ the Jacobi elliptic function. The above equation exhibits persistent oscillations. We are instead interested in the time average of $\overline{x(t)^2}$ which is given by 

\begin{equation}
    \overline{x_v^2}\equiv\lim_{t\to \infty}\overline{x_v(t)^2} = \frac{\sqrt{r^2+2u v^2}-r}{2u}\,.
\end{equation}
Finally, the Gaussian integral over $v$ 
(i.e., the integral $\langle\cdot \rangle_{v}\equiv \int dv \exp(-Dv^2) \cdot/ \int dv \exp(-Dv^2) $) yields 

\begin{equation}
    \left\langle \overline{x_v^2}\right\rangle_v = -\frac{r}{2u}+\frac{U(-\frac{1}{2},0,D r^2/(2u))}{\sqrt{2Du}}\,,
\end{equation}
where $U$ is Tricomi's confluent hypergeometric function. 
Making the dependence of $u$ on $N$ explicit by $u \to u/N$, we find

\begin{equation}
    \overline{{m}^2} \equiv  \frac{1}{N^2}\overline{\langle S_x(t)^2\rangle} =\frac{1}{N} \left\langle \overline{x_v^2}\right\rangle_v = -\frac{r}{2u}+\frac{1}{\sqrt{N}}\frac{U\left(-\frac{1}{2},0,D Nr^2/(2u)\right)}{\sqrt{2Du}}\,.
\end{equation}
Deep in the ordered phase, we must have $\overline{m^2} = m^2$ where the order parameter $m$ is computed from mean field (see \cref{eq:mean-field}). This is indeed the case:  the above equation captures the initial linear dependence of the order parameter on $r$ as one enters the ordered phase, that is, $\overline{m^2} \sim |r| \propto (1-B)$ when $r<0$, while the exact (mean-field) solution within the ordered phase is given by $m^2=B(1-B)$. An expression that properly interpolates between the two extremes can be obtained by multiplying the right hand side of the above equation by a factor of $B$. We also fix the coefficient $u=1/2$ to match~\cref{eq:mean-field} deep in the ordered phase where $r<0$. Making explicit the dependence on  $B$ and $B_c$, we find

\begin{equation}\label{eq:last}
    \overline{m^2}  = \frac{B}{B_c}\left[ \left(1-\frac{B}{B_c}\right)  + \frac{1}{\sqrt{N}}\frac{U\left(-\frac{1}{2},0,D N(1-\frac{B}{B_c})^2\right)}{\sqrt{D}}\right]\,.
\end{equation}
In a numerical calculation of the critical point, the above fit only contains two free parameters: the constant $D$ and the critical point $B_c$. 

In the experiment, it is more convenient to study the maximum of  $\langle S_x^2\rangle/N^2$ in time rather than its stationary value. The above analysis remains valid but with an an overall amplitude, which we shall treat as a free parameter in \cref{eq:last}. In Supplementary~\cref{fig:lmgdynamicalphasediagram}, we show that that above function (up to an overall amplitude $\cal A$) is an excellent fit to the order parameter obtained from the exact numerical simulation, and the extracted value of the critical point is almost identical to the exact value of $B_c=1$. 

\begin{figure}[H]
     \centering
     \includegraphics[height=0.35\textwidth]{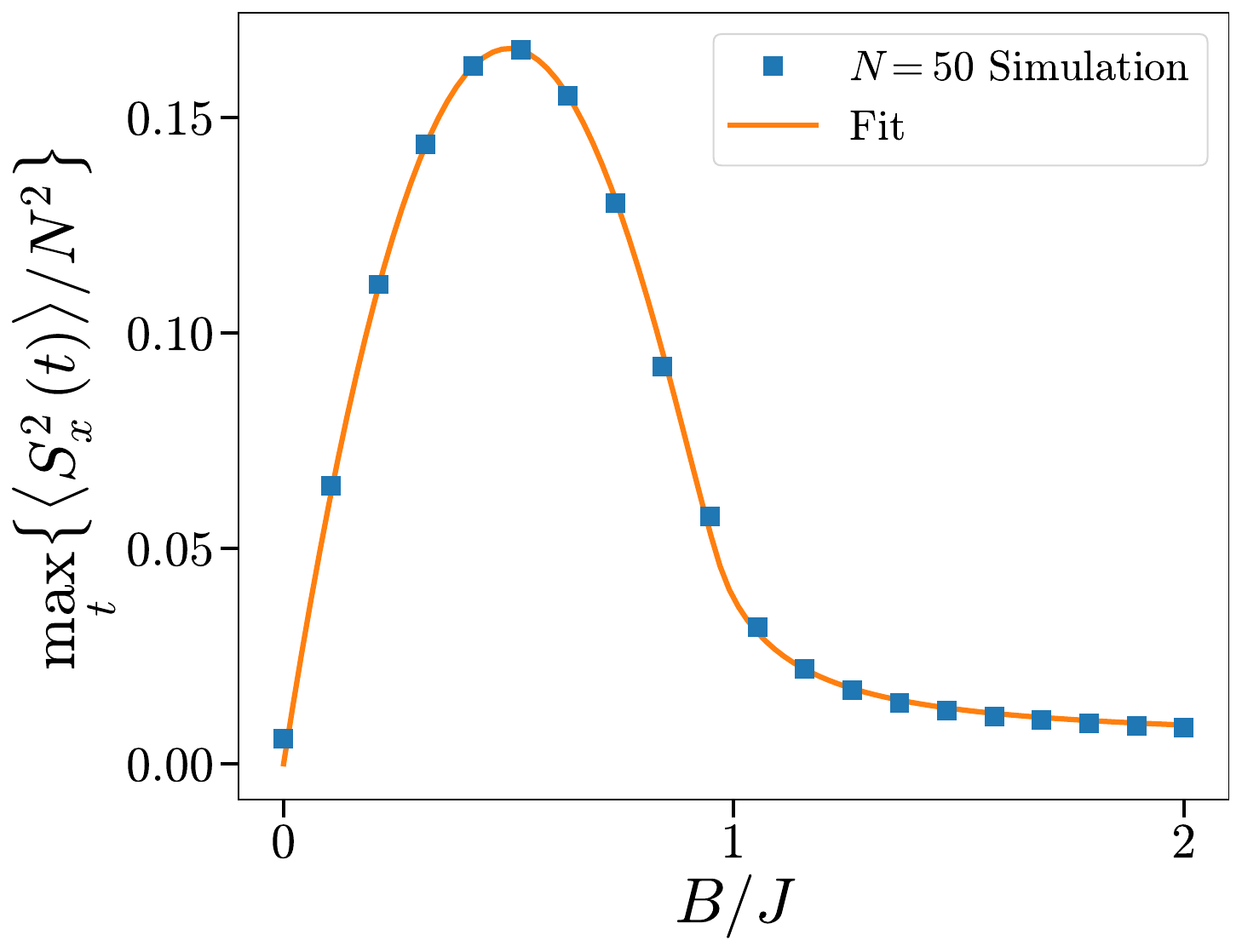}
     \caption{\label{fig:lmgdynamicalphasediagram}The order parameter for the LMG model with $N=50$ extracted from the peak of $\langle S_x^2(t)\rangle$. We also depict the fit function from \cref{eq:last} with an overall factor $\mathcal{A}=16.15(9)$, as well as the fit parameters $D=0.34(1)$ and $B_c=0.976(4)$ in excellent agreement with the exact critical point, $B_c=1$.}
\end{figure}

\subsection{Double (Multiple) Quenches}\label{sec:double_quench}
In this section, we analyze the scaling behaviour of the system as the result of a double quench starting from a disordered initial state to the critical point of the LMG model with $\gamma^x_0=1$ and $\gamma^y_0=0$, and, with a delay, a subsequent quench to the critical point of the LMG model with $\gamma^x_0=0$ and $\gamma^y_0=1$; See Fig.~1(b) of the main text. Both critical points are defined by $B_c=1$. 
Notice that the final Hamiltonian has the Ising interaction along the $y$ direction only, and the dominant fluctuations correspond to the those along the $y$ direction. To parallel the notation in the previous sections, we use the Holstein-Primakoff expansion but now identify, to the lowest order in $1/N$ \cite{Vidal2004, Vidal2005}, 

\begin{equation}
        \sqrt{\frac{2}{N}}S_x \approx p_2\,, \qquad \sqrt{\frac{2}{N}}S_y \approx x_2\,,
\end{equation}
as the dynamical variables upon a second quench, while we reserve $x_1$ and $p_1$ defined similarly to $x$ and $p$ in \cref{eq:HP}, respectively, as the dynamical variables following the first but before the second quench. At the time of the second quench ($t_2=0$), we have  $x_2=p_1$ and $p_2=x_1$ due to continuity. 
The Hamiltonian at each stage is given by $H_i =\frac{1}{2m_i}p_i^2 + \frac{1}{2}m_i\Omega_i^2 x_i^2$ where $i=0,1,2$ correspond to pre-quench, first-quench, and second-quench variables; we also identify $x_0=x_1$ and $p_0=p_1$. To characterize fluctuations at the quadratic order, we shall consider a finite distance from the critical point at all stages of the dynamics (but approaching criticality when considering finite-size scaling). The frequencies $\Omega_i$ are then given by

\begin{equation}
    \Omega_i^2 = 4\left(B_i - J\gx_i\right)\left(B_i - J\gy_i\right)
\end{equation}
and similarly for $m_i$; the exact values of $m_i$ are unimportant and we just remark that they remain finite even as we approach the critical point.

Next, we characterize the dominant fluctuations ($\langle x_2^2\rangle$) at late times from those of $\langle x_1^2\rangle$ and $\langle p_1^2\rangle$ long after the first quench but before the second quench. 
Finally taking the long-time average, we find 

\begin{equation}
    \overline{\expectation{x_2^2}} = \frac{\overline{\expectation{p_1^2}}}{2} + \frac{\overline{\expectation{x_1^2}}}{2m_2^2\Omega_2^2} = \frac{m_0^2\Omega_0^2}{16} + \frac{m_1^2 \Omega_1^2}{16 m_0^2 \Omega_0^2} + \frac{1}{8 m_0 \Omega_0 m_2^2 \Omega_2^2} + \frac{m_0 \Omega_0}{8 m_1^2 \Omega_1^2 m_2^2 \Omega_2^2}.
\end{equation}
where the frequencies $\Omega_i$ are defined above. 
For two successive quenches to the vicinity of critical points, $\Omega_2\ll\Omega_1\ll \Omega_0$, the above expression reduces to

\begin{equation}
    \frac{1}{2}\overline{\expectation{x_2^2}} \sim \frac{m_0\Omega_0}{16 m_1^2\Omega_1^2 m_2^2\Omega_2^2}.
\end{equation}
Using the equipartition theorem $\frac{1}{2}m_2 \Omega_2^2 \langle x_2^2\rangle \sim \frac{1}{2}T $, we can now identify an effective temperature upon the second quench (denoted by subscript 2) as

\begin{equation}\label{eq:T_eff_2}
    T_{\mathrm{eff},2} = \frac{m_0\Omega_0}{8 m_2 m_1^2 \Omega_1^2}.
\end{equation}
Notice, however, that the effective temperature diverges as we tune the first quench to the critical point, $\Omega_1 \to 0$. Such divergence hints at a critical behaviour that is strongly athermal (see also Ref.~\cite{Titum20}). 

To characterize the critical behaviour, we can set up a scaling analysis using semiclassical dynamics. 
Using the continuity of these functions at $t_2=0$ where $x_2=p_1$ and $\dot x_2 \sim p_2=x_1$, the initial conditions for $x_2$ and its time derivative are given  by a distribution function; we can alternatively characterize the same information via the expectation values 

\begin{equation}\label{eq:x2_scaling}
    \langle x_2^2  \rangle \sim N^0  , \quad \langle \dot x_2^2\rangle   \sim N^{\alpha_1} , \qquad  \mbox{at} \,\, t_2=0\,.
\end{equation}
For a uniform notation, we have identified $\alpha_1=\alpha$ as the critical exponent emerging in the first quench. 
To this end, a rather similar scaling analysis gives the partition function 

\begin{equation}\label{eq:Z_delta_fn_2}
    Z=\int Dx_2 \delta[x_2(0)] P\left[(\dot x_{2}(0))^2/N^{\alpha_1}\right] 
    \delta\left(m_2\ddot x_{2}+r_2x_{2}+\frac{u_{2}}{N}x_{2}^3\right)\,,
\end{equation}
where $r_2= m_2\Omega_2^2$ and $u_2=2\gamma^y_2$; we have also defined $x_2(0)\equiv x_2(t_2=0)$ for ease of notation (similarly for $\dot x_2(0)$).
To arrive at this equation, we have used the scaling limit where the quantum vertex can be ignored and the fluctuations of $x_2$ right after the second quench ($t_2=0$) becomes a delta function. 
On the other hand, we keep track of the  initial fluctuations of $\dot x_2$ and its scaling with system size $N$ via the the distribution function $P$;
for our purposes, only the scaling behaviour, and not the exact form of the function $P$, is important. We stress that the scaling variable in the argument of the function $P$ is dictated by scaling behaviour in \cref{eq:x2_scaling} at the onset of the second quench.

Next, we observe that the partition function is scale-invariant (at $r_2=0$) upon the transformation

\begin{equation}
    t\to t/ \lambda, \quad x \to x/\lambda^{\frac{1+\alpha_1}{1-\alpha_1}}, \quad N \to  N/\lambda^{\frac{4}{1-\alpha_1}}\,.
\end{equation}
We can then immediately write a scaling relation for fluctuations 

\begin{equation}
    \langle x_2(t_2)^2\rangle = N^{\alpha_2} g(t_2/N^{\zeta_2})\,,
\end{equation}
with $g$ a scaling function and the critical exponents identified as

\begin{equation}
    \alpha_2 = \frac{1+\alpha_1}{2}, \qquad \zeta_2 = \frac{1-\alpha_1}{2}\,.
\end{equation}
Using $\alpha_1=1/2$ in the first quench, we find $\alpha_2=\frac{3}{4}$  and $\zeta_2=\frac{1}{8}$. These predictions are in excellent agreement the exact numerical simulation shown in Supplementary~\cref{fig:scaling}(c).  

At this point, we can also determine the scaling of the effective temperature. Note that \cref{eq:Z_delta_fn_2} remains scale-invariant even for nonzero $r_2$ upon scaling $r_2 \to \lambda^2 r_2$, which then determines the scaling behaviour of this variable as $r_2\sim N^{-2\zeta_2}$. Now recalling the definition of the effective temperature $\frac{1}{2}\Omega_2^2 \overline{\langle x_2^2\rangle} \sim \frac{1}{2} T_{\rm eff,2}$ together with the facts $r_2 \propto \Omega_2^2 \sim N^{-2\zeta_2}$ and $\langle x_2^2\rangle \sim N^{\alpha_2}$, we find

\begin{equation}
\label{eq:Teff_double}
    T_{\rm eff,2}
    = N^{\alpha_2-2\zeta_2}\,.
\end{equation}  
The same relation holds for the first quench by changing all the subscripts to 1. While in the latter case, $\alpha_1 -2\zeta_1 =0$ indicating a constant effective temperature, for a double quench $\alpha_2 - 2\zeta_2 = 1/2$ highlighting a nontrivial scaling of the effective temperature and a genuinely nonthermal critical behaviour. 

We also remark that more generally for $k+1$ consecutive critical quenches $\gamma^x\to\gamma^y \to\gamma^x \to \cdots$ (showing only the nonzero anisotropy parameter) starting from a disordered state, we find the non-equilibrium critical exponents:

\begin{equation}\label{eq:iterative}
    \alpha_{k+1}= \frac{1+\alpha_k}{2}, \qquad \zeta_{k+1}= \frac{1-\alpha_k}{4}\,,
\end{equation}
and the effective temperature scaling

\begin{equation}\label{eq:Teffff}
    T_{\mathrm{eff},k} \sim N^{\alpha_k-2\zeta_k}\,,
\end{equation}
with $\alpha_0=0$. One can solve the above iterative equations to find the exponents 

\begin{equation}
    \alpha_k = 1- 2^{-k}, \qquad \zeta_k = 2^{-k-1}\,.
\end{equation}
The critical exponents for single and double quench are recovered by setting $k=1$ and 2, respectively. 
We can also determine the critical scaling of the temperature from \cref{eq:Teffff} as

\begin{equation}
  T_{\mathrm{eff},k} \sim N^{1-\frac{1}{2^{k-1}}}\,.
\end{equation}

\section{Power-law Interacting Spin Chain}\label{sec:long-range}
In this section, we extend our analysis to include the long-range model in \cref{eq:H_generalized} where the interaction $J_{ij}$ falls off as a power-law with the distance, $J_{ij}\sim 1/|i-j|^p$. 
We show, using a spin-wave analysis, that the universal behaviour discussed in the text is governed by a collective mode as long as $0<p<1$, hence the same universality class as the LMG model discussed in the previous section. 
Furthermore, we show that the critical point is consistent with the mean-field (or, the LMG model's) prediction, i.e., $B_c = 1$, in the thermodynamic limit.

We begin by applying the Holstein-Primakoff transformation to \cref{eq:H_generalized} at the level of individual spins

\begin{align}
\sigma^{-}_i = (\sigma_i^+)^\dagger= \sqrt{1 - a^\dagger_i a_i} a_i \approx a_i,\qquad 
\sigma^z_i = 1 - 2 a^\dagger_i a_i\,,
\end{align}
where we have neglected the non-linear terms. This approximation is valid as long as we remain in the disordered phase where the total spin is fully polarized along the $z$-direction, $\langle S_z \rangle  = N/2$. Excitations of the bosonic degrees of freedom amount to fluctuations beyond this state.
Under this approximation, the bosonic form of the Hamiltonian is given by

\begin{equation}
    H =  -\frac{1}{2\mathcal{J}}\sum_{ij} J_{ij} (a_i + a^\dagger_i) (a_j + a^\dagger_j) + 2B \sum_i a^\dagger_i a_i\,,
\end{equation}
where we have neglected an unimportant constant, and have only included the quadratic terms in the Hamiltonian. 
This Hamiltonian, being quadratic in the bosonic operators, can be diagonalized via a Bogoliubov transformation. The form of this transformation depends on the form of $J_{ij}$. We first consider power-law interactions with periodic boundary conditions where simple analytical expressions can be obtained, and then extend our results to open boundary conditions. In the next section, we apply the same analysis to the experimental interaction matrix and contrast the results against those with a purely power-law interaction.

\subsection{Power-Law $J_{ij}$ with Periodic Boundary Conditions}
For illustrative purposes, we first consider the case where $J_{ij}$ is described by a power-law  subject to periodic boundary conditions:

\begin{equation}
    J_{ij} = \frac{J}{r_{ij}^p}, \qquad \mbox{with} \quad r_{ij} = \text{min}(|i-j|,N-|i-j|) \,,
\end{equation}
and $J_{ij} = 0$ when $i=j$.
We can bring the Hamiltonian into a convenient form by going to the Fourier basis,

\begin{equation}
    a_j = \frac{1}{\sqrt{N}}\sum_k e^{i j k} a_k\,,
\end{equation}
where $k = 2\pi n/N$, $n \in \left\{0, \ldots, N-1\right\}$. The Hamiltonian now becomes diagonal in $k$,

\begin{equation}\label{eq:spinwavehamiltonian}
    H =  \sum_k \begin{pmatrix} a^\dagger_k \\ a_{-k}\end{pmatrix}^T \begin{pmatrix} -\frac{\tilde J_k}{2\mathcal{J}} + B  & -\frac{\tilde J_k}{2\mathcal{J}} \\ -\frac{\tilde J_k}{2\mathcal{J}} & -\frac{\tilde J_k}{2\mathcal{J}} + B \end{pmatrix}\begin{pmatrix} a_k \\ a^\dagger_{-k}\end{pmatrix}\,,
\end{equation}
but remains to be diagonalized in terms of the bosonic operators. Here, we have defined 

\begin{equation}
    \tilde J_k = \sum_{r = -(N-1)/2}^{(N-1)/2} J_{r} e^{-irk} = 2J \sum_{r = 1}^{(N-1)/2} \frac{1}{r^p}\cos(rk)\,,
\end{equation}
and assumed that $N$ is odd for simplicity. Next, we employ the Bogoliubov transformation,

\begin{equation}
    a_k = \cosh(\theta_k) b_k + \sinh(\theta_k) b^\dagger_{-k}\,,
\end{equation}
where $\theta_{k} = \theta_{-k}$, and the coefficients have been chosen to ensure that the new operators $b_k$ satisfy the bosonic commutation relation $[b_k, b^\dagger_{k'}] = \delta_{kk'}$. To determine $\theta_k$, we demand that this transformation diagonalizes \cref{eq:spinwavehamiltonian} as $H = \sum_k \omega_k b^\dagger_k b_k$, which leads to the relation

\begin{equation}
    \tanh(2\theta_k) = \frac{\tilde J_k /2\mathcal{J}}{B - \tilde J_k/2\mathcal{J}}\,,
\end{equation}
determining $\theta_k$, as well as the mode frequencies

\begin{equation}
    \omega_k= \sqrt{B(B -  \tilde J_k/\mathcal{J})}\,.
\end{equation}
With $\tilde J_0/\mathcal{J} = (N-1)/N$, we have $\omega_0 = \sqrt{B(B- 1)}$ in the thermodynamic $N\to \infty$ limit. 
Therefore, the collective mode corresponding to $k=0$ becomes gapless (or, \textit{softens}) at the phase transition, and should be identified as the soft mode of the phase transition. For $B < 1$, the dispersion becomes complex valued and the Holstein-Primakoff transformation is no longer valid. 
Most importantly, we find that all the other modes beside the $k=0$ mode remain gapped for long-range interactions with $p<1$.  
This is because $\tilde J_k / {\cal J}< 1$ for any $k>0$, therefore $\omega_k > 0$ at or away from the phase transition within the normal phase, $B \geq 1$. Indeed, one can show that the gap between the $k=0$ mode and the next mode corresponding to $k=2\pi/N$ remains finite for $p<1$ even in the thermodynamic limit $N\to \infty$ \cite{Defenu2021}:

\begin{equation}
    \lim_{N\to \infty} \omega_{k=2\pi/N} - \omega_{k=0} > 0, 
\end{equation}
at or beyond the critical point.  
This implies that the critical behaviour of long-range interactions with $p<1$ is in the same universality class of the LMG model. 

\subsection{Power-Law $J_{ij}$ with Open Boundary Conditions}\label{sec:open_boundary}
It is generally expected that the bulk properties of a large system would be insensitive to boundary conditions at least for short-range interactions. The role of boundaries are, however, more pronounced in the presence of long-range interactions. Specifically, for truly long-range models with $p<1$, the boundary conditions can even change the qualitative behaviour \cite{Hauke2013}. In this section, we show that open boundary conditions do not alter our conclusions for a long-range interacting chain with periodic boundary conditions.
To perform a spin-wave analysis for $J_{ij}$ (again taking $J_{ii}=0$) without translation invariance, we cannot use the Fourier basis and instead carry out the Bogoliubov transformation directly in real space. We rewrite the Hamiltonian in the more convenient form

\begin{equation}\label{eq:OBChamiltonian}
    H = \sum_{i,j}\begin{pmatrix}
        a_i \\ a^\dagger_i
    \end{pmatrix}^\dagger \begin{pmatrix}
        A_{ij} & C_{ij} \\ C_{ij} & A_{ij}
    \end{pmatrix}\begin{pmatrix}
        a_j \\ a^\dagger_j
    \end{pmatrix}\,,
\end{equation}
where $A_{ij} = -J_{ij}/2\mathcal{J} + B\delta_{ij}$, $C_{ij} = -J_{ij}/2\mathcal{J}$. Using the fact that the matrices $\mathbf A$ and $\mathbf C$ commute (the bold notation denotes matrix or vector objects), we can directly determine the Bogoliubov transformation which brings the above Hamiltonian into a diagonal form $H = \sum_k \Lambda_k b^\dagger_k b_k$  \cite{colpa_diagonalization_1978}. Specifically, the mode energies $\Lambda_k$ are given by the square root of the eigenvalues of the matrix

\begin{equation}
    \mathbf M = (\mathbf A+\mathbf C)(\mathbf A- \mathbf C) = \mathbf A^2 - \mathbf C^2 \,.
\end{equation}
The eigenvalues can be determined numerically for an arbitrary interaction matrix $J_{ij}$.\par
In Supplementary~\cref{fig:spinwave1}, we consider $J_{ij}$ matrix describing a power-law interacting chain with open boundary conditions and plot the mode frequencies $\Lambda_{0}^2$ and $\Lambda_{1}^2$ for $N = 50$ and $p = 0.9$; the value of $p$ is chosen the mirror the experimental model although the latter features an exponential decay at large distances (see the next section). $\Lambda_0$ corresponds to the soft mode which, by definition, becomes gapless at the phase transition, $B \simeq 1$. (We recover $B_c= 1$ exactly in the thermodynamic limit.) In addition, we see that $\Lambda_1$ remains gapped even at the critical point, and therefore does not contribute the critical behaviour. Again, $\Lambda_0^2$ becomes negative for $B < 1$, and our bosonic mapping is no longer valid.
\begin{figure}[H]
    \centering
         \subfloat{\includegraphics[width = 0.48\textwidth]{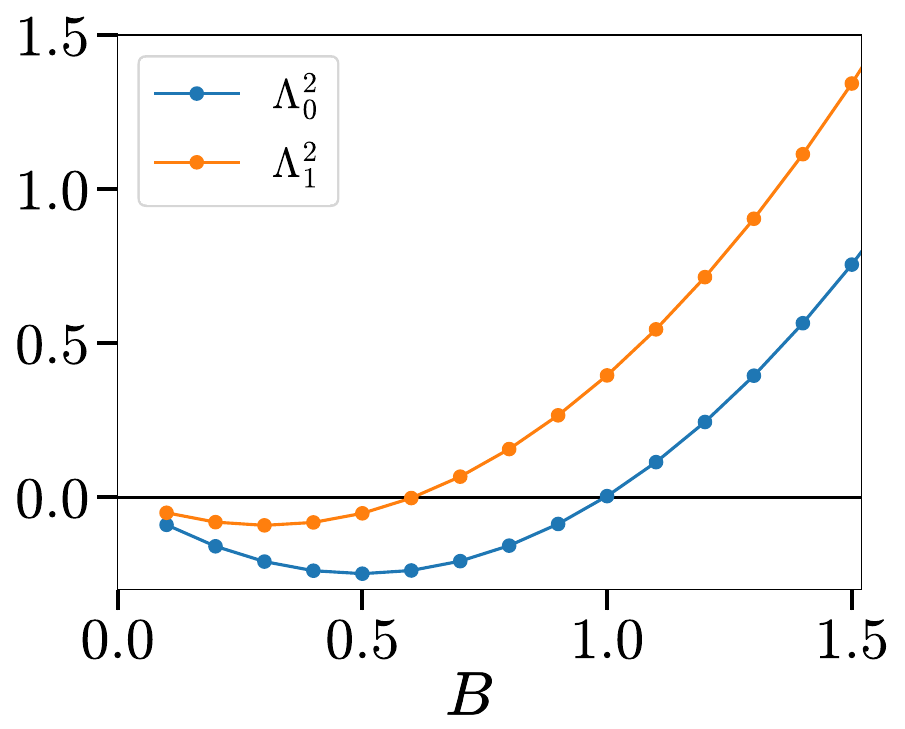}}
    \caption{The two lowest energy eigenvalues $\Lambda_0$ and $\Lambda_1$ corresponding to a power-law interaction with open boundary conditions,  $p = .9$ for a system size of $N = 50$. While the lowest-frequency mode softens (i.e., $\Lambda_0 \to 0$) at the critical point, all the other modes remain gapped.
    }
    \label{fig:spinwave1}
\end{figure}

\section{Long-range Interacting Spin Chain with Experimental $J_{ij}$}\label{sec:experiment_long-range}
In practice, the experimental values of the long-range coupling are not completely described by a power-law decay, and at large distances decay exponentially; see \cref{sec:Jij comparison}. In spite of this, we show that, based on a spin-wave analysis, a phase transition still occurs close to the predicted critical point and that a single soft mode is responsible for the critical behaviour. To this end, we first note that the experimental Hamiltonian is given by \cref{eq:H_generalized} without the Kac factor. In plotting various quantities however, we divide by the corresponding Kac factor to compare and contrast different cases on equal footing.
\begin{figure}[H]
    \centering\subfloat{\includegraphics[width=0.5\linewidth]{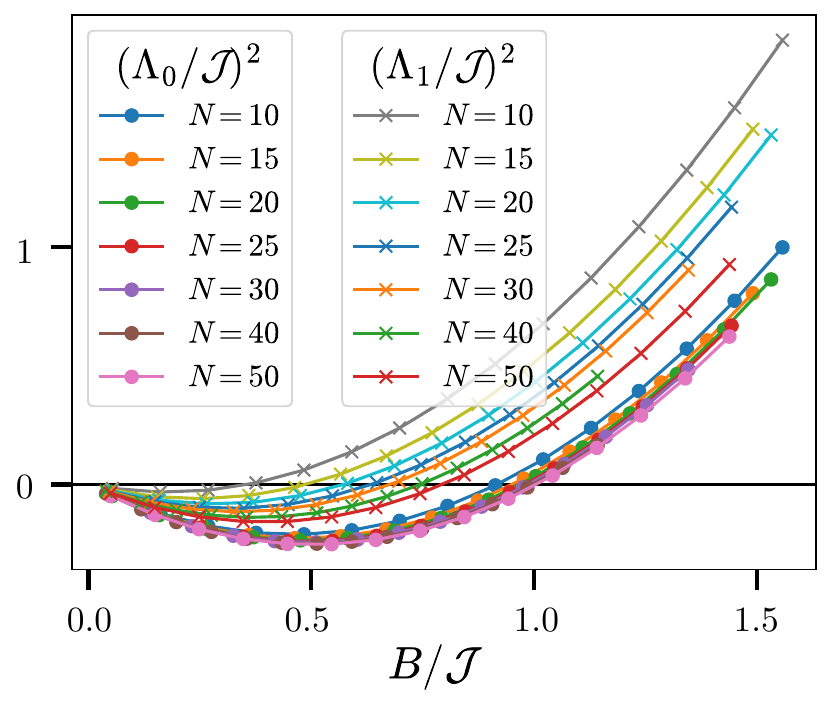}}
    \hfill 
    \caption{The two lowest energy eigenvalues $\Lambda_0$ and $\Lambda_1$ versus the (Kac-normalized) transverse field $B/\mathcal{J}$, corresponding to the experimental $J_{ij}$ for different system sizes; $\mathcal{J}$ is the Kac factor. The lowest-frequency mode softens, $\Lambda_0 \to 0$, approximately at the mean-field predicted critical point $B_c/\mathcal{J} = 1$. While the next eigenvalue $\Lambda_1$ becomes smaller for larger and larger system sizes, in contrast with a purely power-law $J_{ij}$, it remains finite even at the critical point (specifically, $\Lambda_1 /{\cal J} \approx 0.5$ for $N=50$ at $B=B_c$).
    }
    \label{fig:spinwavegap}
\end{figure}
In Supplementary~\cref{fig:spinwavegap}, we plot the first two lowest energy eigenvalues as a function of $B$ for different system sizes each with the corresponding experimental values of $J_{ij}$; we have divided $B$ by the Kac factor $\mathcal{J}$ on the $x$-axis and re-scaled the eigenvalues accordingly. The points where the lowest-frequency mode softens (i.e., $\Lambda_0 \to0)$ agree very well with the mean-field prediction $B_c/\mathcal{J} \sim 1$. However, the gap between $\Lambda_1$ and $\Lambda_0$ (in units of the Kac factor) exhibits a strong dependence on system size, in contrast to the power-law case. This suggests a departure from the collective nature of the power-law interactions with $p < 1$ near the critical point. As $\Lambda_1$ becomes smaller, the corresponding mode becomes populated as well and contributes to the correlations. At the same time, we note that  the latter mode still remains gapped even for $N=50$ where $\Lambda_1 /{\cal J} \sim 0.5$ at the critical point $B/{\cal J} \approx 1$ while $\Lambda_0 =0$. Therefore, the soft mode corresponding to $\Lambda_0$ should govern the critical behaviour despite the model's departure from infinite- or truly long-range models considered before.

For a quantitative estimate of the contribution due to the first gapped mode (corresponding to $\Lambda_1$), let us first consider a toy model where a harmonic oscillator in its ground state undergoes a sudden frequency quench $\Omega_0\to\Omega$. This quench excited the harmonic oscillator and creates a population of $n+1/2 = \frac{1}{4}(\Omega/\Omega_0+ \Omega_0/\Omega)$. Now considering the first gapped mode, we can substitute $\Omega_0 = 2B$ and $\Omega = \Lambda_1$, which gives 

\begin{equation}
    n_1  
    \sim \frac{1}{2\Lambda_1/ {\cal J}} \sim 1\,,
\end{equation}
where the numerical estimate is computed by setting $B = B_c = {\cal J}$ and 
 $\Lambda_1/{\cal J} \sim 0.5$ consistent with the data for $N=50$ in Supplementary~\cref{fig:spinwavegap}. 
This should be contrasted against the soft mode whose population scales with system size, according to our scaling analysis, as $n_0 \sim \sqrt{N}$ at the critical point. For $N=50$, for example, this means that the soft mode is populated by $n_0 \lesssim 10$ in contrast with $n_1 \sim 1$.
While the higher excited modes may be populated with $n\sim 1$, the soft mode still governs the critical behaviour at the phase transition.

\section{Experiment vs Theory}\label{sec:expt_vs_theory}
In this section, we contrast various theoretical predictions against the experimental results. 
\subsection{Finite-Size Behavior of the Order Parameter}
Mean-field analysis, together with finite-size corrections, yields \cref{eq:last} up to an overall coefficient. Partial data and the comparison against the experiment is shown in Fig.~2(b) of the main text. In Supplementary~\cref{fig:finite_size_mean_field}, we provide additional information for larger system sizes and report the best fit for the critical point $B_c$. We find that the experimental data are in good agreement with the mean-field prediction $B_c/{\cal J} = 1$.
\begin{figure}[H]
    \centering
    \includegraphics[width=0.48\textwidth]{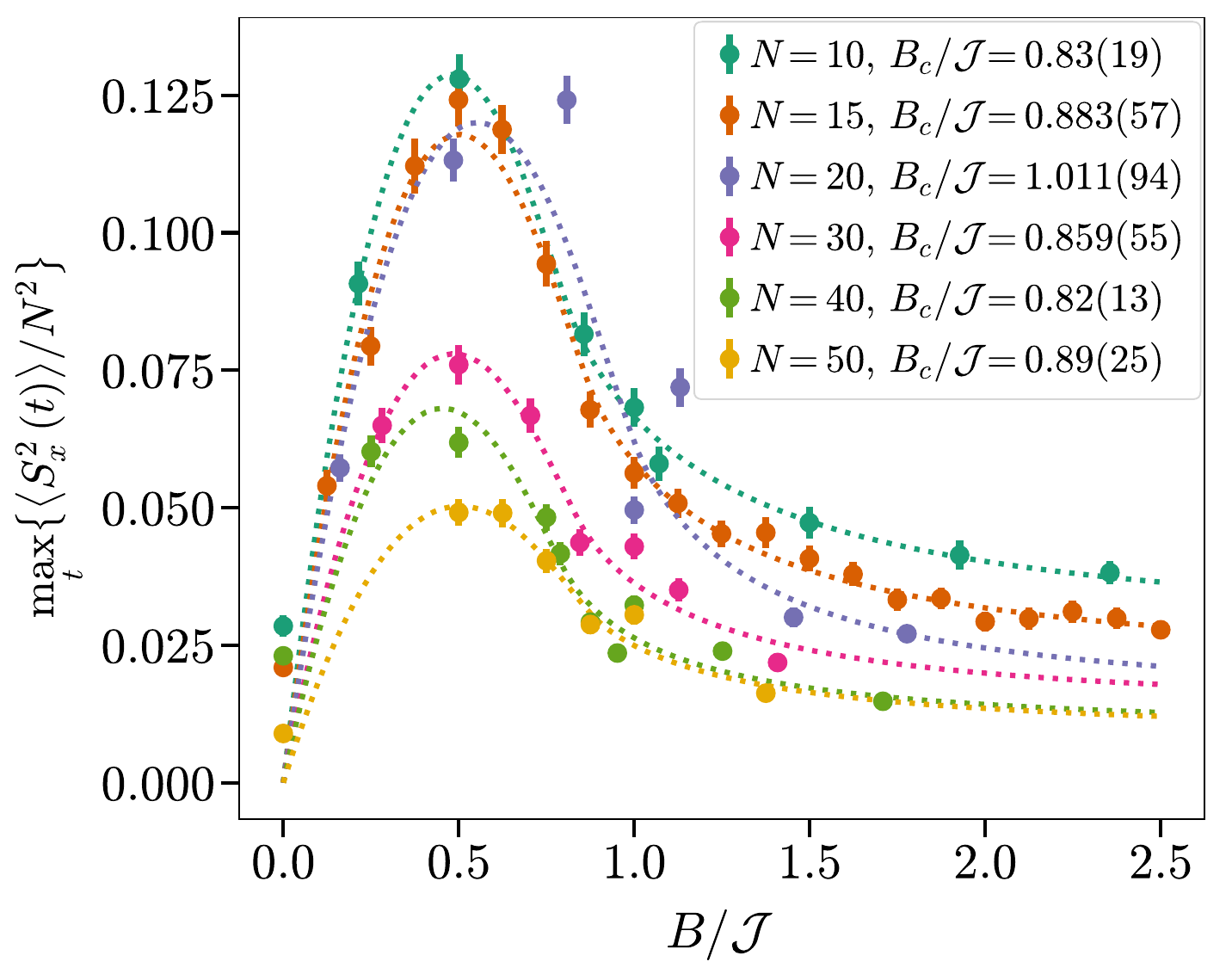}
    \caption{Experimental data for the order parameter against the finite-size corrected mean-field prediction in \cref{eq:last}; $\cal J$ is the Kac normalization. The experimental data are well described by the theoretical finite-size correction.}\label{fig:finite_size_mean_field}
\end{figure}

\subsection{Long-Range Character of $J_{ij}$ Matrix}\label{sec:Jij comparison}
In Supplementary~\cref{fig:jijcomparison}, we present the numerically calculated average interaction $J(r)=\frac{1}{N-r}\sum_i J_{i,i+r}$ as a function of distance $r$ for $N=10$, and $50$. We first fit the numerical values to a pure power-law $J(r)\sim J(1)/r^p$ (dashed line in Fig. \ref{fig:jijcomparison}) and find the exponent $p$ to be around $0.9$ for both system sizes.

\begin{figure}[H]
    \centering
    \includegraphics[width=0.5\textwidth]{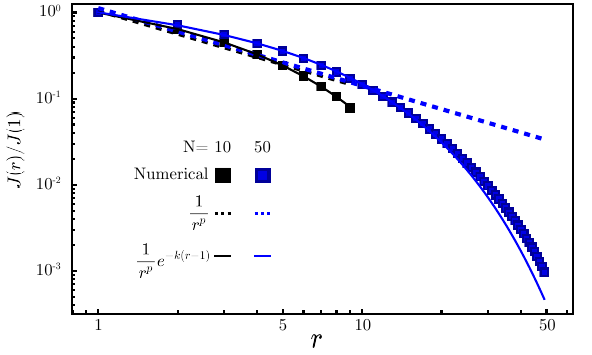}\caption{\label{fig:jijcomparison}Interaction profile $J(r)$ vs $r$. The numerical values of $J(r)$ are shown for system sizes $N=10,\: 50$ (see the squares). A power-law fit, $J(r)/ J(1)\sim  r^{-p}$, gives $p=0.89$ for both system sizes (see the dashed lines), and deviates from $J(r)$ at large distances. A better fit is provided by the product of a power-law with an exponential decay, $J(r)/ J(1)\sim r^{-p} e^{-k(r-1)}$, and gives $p=0.306,\: k=0.231$ for $N=10$, and $p=0.306,\: k=0.135$ for $N=50$ (see the solid lines). The latter fit captures the trend of $J(r)$ at larger distances as well.}
\end{figure}

However, we note that the power-law fit deviates quite significantly from the calculated values, specifically at large distances and for larger system sizes. We further fit $J(r)$ against the product of a power-law and an exponentially decaying function of the form $J(r)=\frac{J(1)}{r^p}e^{-k(r-1)}$ \cite{Pagano2020}; see the solid lines in Supplementary~\cref{fig:jijcomparison}. We find the exponents $p=0.306,\: k=0.231$ for $N=10$ and $p=0.306,\: k=0.135$ for $N=50$ and note that the fit represents the numerical values closely. The similarity of the power-law exponents for two system sizes indicates that the experimental interaction matrices are self-similar at short distances. In ~\cref{sec:correlation_matrix}, we show that the spatial correlations after a single quench shows a slower decay compared to the spin-spin interactions. This behaviour is consistent with the critical correlations building up in the system.

\subsection{Correlation Matrix}\label{sec:correlation_matrix}
In this section, we provide the correlation matrix $\expectation{\sigma_i^x\sigma_j^x}$ after a single quench. Specifically, we consider the average correlations between sites at a fixed distance $r$ and at the same time as the peak of $\Sxsq$, defined as ${C(r) = \frac{1}{N-r}\sum_j \expectation{\sigma_j^x \sigma^x_{j+r}}}$. In Supplementary~\cref{fig:corr}, we show the correlations in the experiment and contrast them against the exact simulation of the experimental model when available. Supplementary~\cref{fig:corr}(a) shows that the correlation function is in reasonable agreement with the exact simulation. 
Sources of the remaining discrepancy are expected to include imperfect experimental detection and decoherence effects. 

Supplementary~\cref{fig:corr}(b) shows the correlation function for a system of size  $N=50$. Again, we find a much slower decay of the correlation function compared to the interaction profile $J(r)$, indicating criticality and long-range correlations in the system. On the other hand, the decay of the correlation function is relatively pronounced, a feature that should be contrasted against the LMG model (see \cref{sec:LMG}) which shows no decay due to infinite-range interactions, and the purely power-law models (see \cref{sec:long-range}) which are expected to show a relatively slow decay of the correlations with distance. This observation once again underscores the departure of the experiment from  infinite- or truly long-range models with $p<1$. In light of this, it is quite remarkable that the critical behaviour observed in the experiment is almost identical to the latter models (cf. \cref{sec:experiment_long-range}).
\begin{figure}[H]
    \centering
    \includegraphics[width=0.8\textwidth]{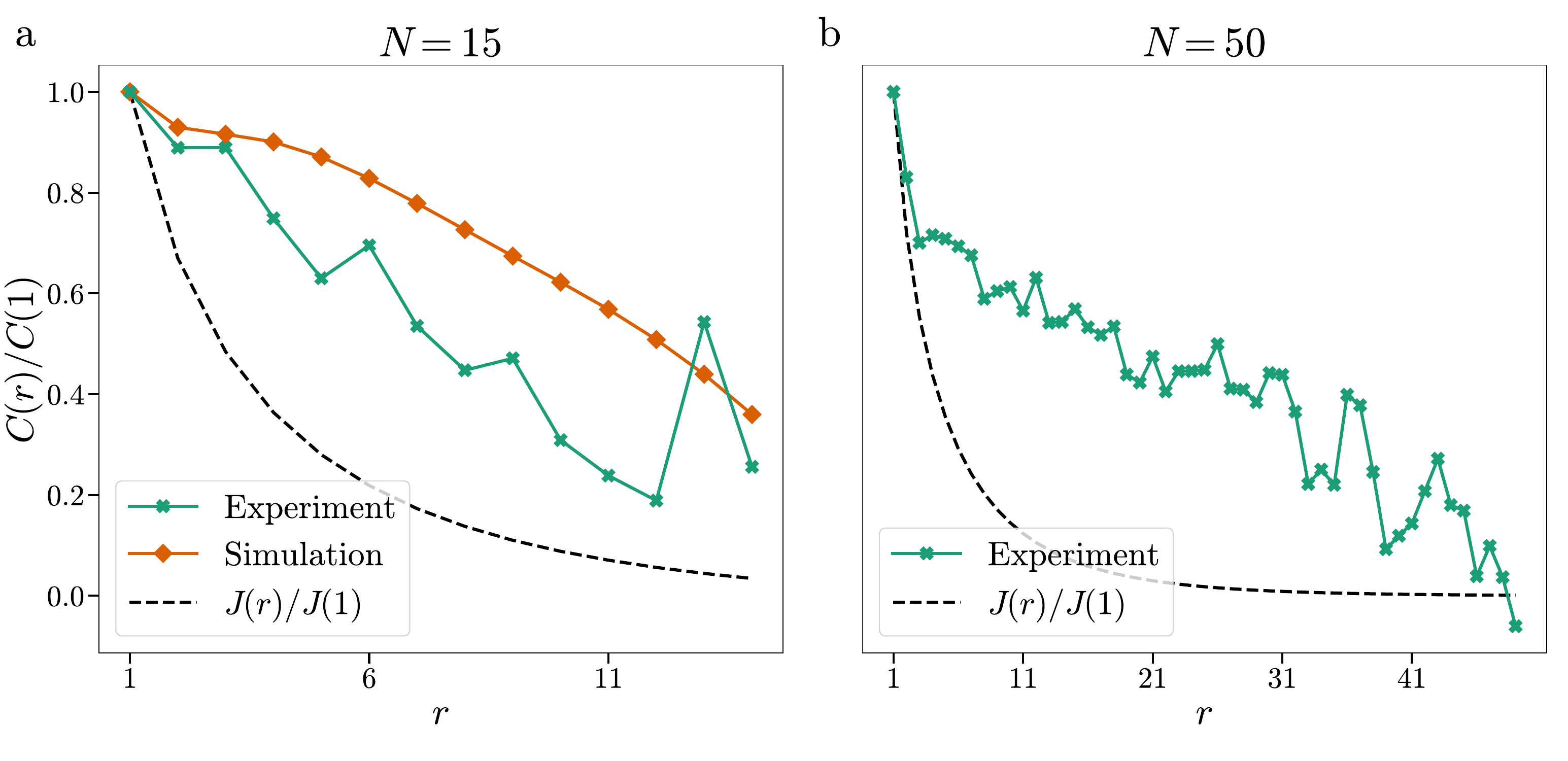}
    \caption{\label{fig:corr}Correlations upon a sudden quench to the critical point for (a) $N=15$ and (b) $N=50$; these correlations are measured/computed at the same time as the peak of $\Sxsq$. (a) 
    Correlations decay slowly compared to $J(r)$ indicating the build-up of critical fluctuations. 
    (b) The data for $N=50$ shows consistent results where correlations decay rather slowly compared to the interaction $J(r)$, indicating critical correlations. 
    On the other hand, correlations decay significantly with $r$ in contrast with the infinite- or truly long-range chain chains, reflecting the distinct character of the experimental system (see \cref{sec:Jij comparison}).
    }
\end{figure}

\subsection{Double Quench Switch Times}\label{sec:switch_times}

Our experimental and simulation data for the critical-to-critical quench does not collapse perfectly, even with exponents found by optimizing our objective function. We have found this to be primarily due to deviations in the time at which the second quench was performed in the experimental system, which were then used in the simulations.

The second quench should be performed at the same characteristic time for all system sizes; that is, the same critically scaled time of the first quench, $\mathcal{J} t/N^{\zeta_1}$ where $\zeta_1 \approx 1/4$. However, in order to avoid using the expected results in the data analysis, in Figs. 4 and 8 of the main text we restricted ourselves to choosing the switch time for each dataset (whether experiment or simulation) as the time at which the fluctuations in the experimental dataset were observed to reach a maximum, which has an inherent uncertainty due to the limited data sampling and experimental error sources. As a result, the switch times deviated around this value between system sizes.
\begin{figure}[H]
    \centering
    \includegraphics[width=0.9\linewidth]{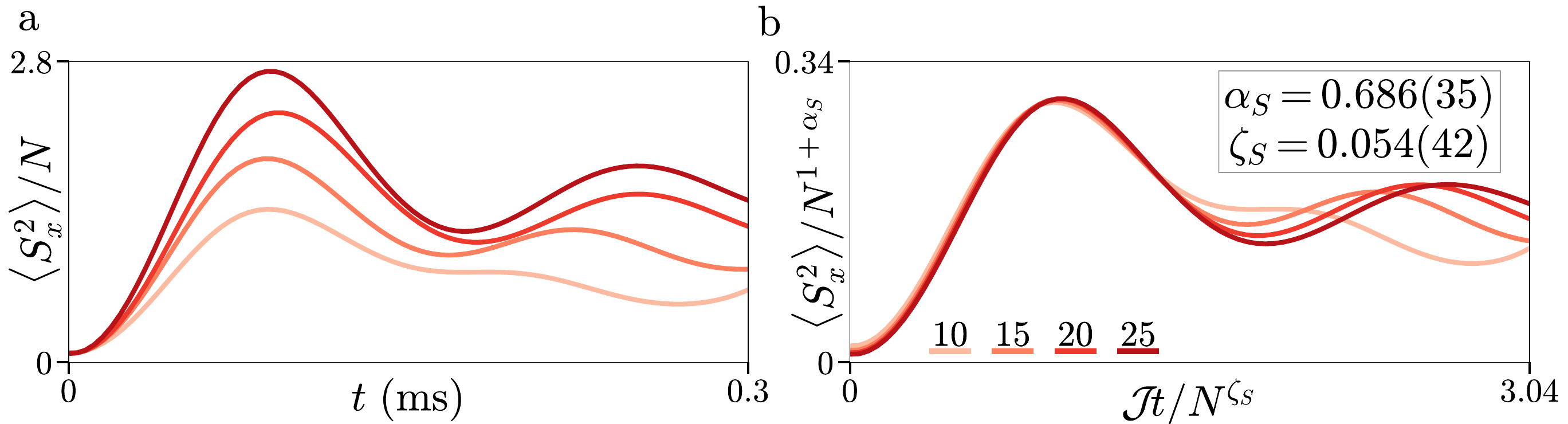}
    \caption{\label{fig:perfect_switch_times}Simulation of the experimental critical-to-critical quench using corrected switch times. We report the unscaled fluctuations in (a) and scaled fluctuations in (b). These are to be compared to Fig. 8 a,b of the main text, which instead use the best experimental estimates for these switch times. The critical exponents in (b) are calculated using the dynamics up to the first minimum after the first maximum, just as in the main text. The collapse up to and including the first peak is nearly perfect with the critical exponents that are in much better agreement with the theory prediction.}
\end{figure}

In Supplementary~\cref{fig:perfect_switch_times}, we show results from simulating the experimental system with identical characteristic times. This figure demonstrates that, with the corrected quench times, the data collapse nearly exactly.

\section{Optimal Collapse Algorithm}\label{sec:optimal_collapse}

To determine the scaled data shown in Figs. 3, 4, 7 and 8 of the main text, we construct an objective function to measure the quality of the collapse of several curves each with some uncertainty. Previous work has used an objective function similar to the weighted-squared-difference, where all curves are linearly interpolated together to produce an estimate for the so-called ``master-curve,'' from which the error is calculated as the distance to each curve~\cite{Houdayer04}. Here we propose an alternative objective function that is more general and less sensitive to noise in the input data.

\begin{figure}[h!]
    \centering
    \includegraphics[width=0.4\linewidth]{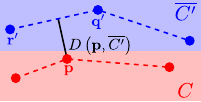}
    \caption{\label{fig:loss_example} Illustration of the calculations involved in the loss function, \cref{eq:loss}.}
\end{figure}

We define a curve in this context as a set of data points with associated uncertainties. Each point is denoted by $\mathbf{p}\equiv [x, y, \Delta x, \Delta y] \equiv [\mbf{p}_x, \mbf{p}_y, \Delta\mbf{p}_x, \Delta\mbf{p}_y]$, and a curve with $k$ data points by $C\equiv\{\mathbf{p}\}\in\mathbb{R}^{4\times k}$. We first consider the case of two curves, $C$ and $C^\prime$ with $k$ and $k^\prime$ data points with $k$ and $k^\prime$ not necessarily equal. We begin by equally normalizing the two curves in both $x$ and $y$ such that they are wholly contained within the region $[0,1]\times[0,1]$. Denote these normalized curves by $\overline{C}$ and $\overline{C^\prime}$ given by

\begin{equation}
    \begin{aligned}[c]
        \begin{aligned}[c]
            Y_{\mathrm{max}} &= \max_{\mbf{p}\in C\cup C^\prime} \mbf{p}_y,\\
            Y_{\mathrm{min}} &= \min_{\mbf{p}\in C\cup C^\prime} \mbf{p}_y,
        \end{aligned}
        \qquad&
        \begin{aligned}[c]
            X_{\mathrm{max}} &= \max_{\mbf{p}\in C\cup C^\prime} \mbf{p}_x,\\
            X_{\mathrm{min}} &= \min_{\mbf{p}\in C\cup C^\prime} \mbf{p}_x,
        \end{aligned}\\
        \overline{C} = \left\{ \frac{\mbf{p}-X_{\mathrm{min}}}{X_{\mathrm{max}}-X_{\mathrm{min}}}, \frac{\mbf{p}_y-Y_{\mathrm{min}}}{Y_{\mathrm{max}}-Y_{\mathrm{min}}},\right.&\left.\frac{\Delta\mbf{p}_x}{X_{\mathrm{max}} - X_{\mathrm{min}}}, \frac{\Delta\mbf{p}_y}{Y_{\mathrm{max}}-Y_{\mathrm{min}}}\middle|\mbf{p}\in C\right\},\\
        \overline{C^\prime} = \left\{ \frac{\mbf{p^\prime}-X_{\mathrm{min}}}{X_{\mathrm{max}}-X_{\mathrm{min}}}, \frac{\mbf{p^\prime}_y-Y_{\mathrm{min}}}{Y_{\mathrm{max}}-Y_{\mathrm{min}}},\right.&\left.\frac{\Delta\mbf{p^\prime}_x}{X_{\mathrm{max}} - X_{\mathrm{min}}}, \frac{\Delta\mbf{p^\prime}_y}{Y_{\mathrm{max}}-Y_{\mathrm{min}}}\middle|\mbf{p^\prime}\in C^\prime\right\}.
    \end{aligned}
\end{equation}
We define the loss function for a single point in $\overline{C}$ relative to $\overline{C^\prime}$, $L\left(\mbf{p} \in \overline{C}, \overline{C^\prime}\right)$, using the square of the distance from the point to the nearest line segment in $\overline{C^\prime}$, weighted by the square of the uncertainty in that distance. If the point in $\overline{C}$ falls outside of the bounds of $\overline{C^\prime}$ then the loss is $0$.

\begin{equation}
    \label{eq:loss}
    L\left(\mbf{p}\in\overline{C}, \overline{C^\prime}\right) =
    \begin{cases}
        0 & \text{if } \mbf{p}_x < \min\limits_{\mbf{p^\prime}\in C^\prime} \mbf{p^\prime}_x \text{ or } \mbf{p}_x > \max\limits_{\mbf{p^\prime}\in C^\prime} \mbf{p^\prime}_x,\\
        D^2\left(\mbf{p}, \overline{C^\prime}\right) / \Delta^2\left(\mbf{p}, \overline{C^\prime}\right) & \text{otherwise},
    \end{cases}
\end{equation}
where $D^2\left(\mbf{p}, \overline{C^\prime}\right)$ is the square of the distance from $\mbf{p}$ to the nearest line segment in $\overline{C^\prime}$, 

\begin{equation}
    \label{eq:D_def}
    \begin{aligned}
        D^2\left(\mathbf{p}, \overline{C^\prime}\right) &\equiv \frac{\left[ \left(\mbf{q^\prime}_x-\mbf{r^\prime}_x\right)\left(\mbf{r^\prime}_y - \mbf{p}_y\right) - \left(\mbf{r^\prime}_x - \mbf{p}_x\right)\left(\mbf{q^\prime}_y - \mbf{r^\prime}_y\right)\right]^2}{\left(\mbf{q^\prime}_x-\mbf{r^\prime}_x\right)^2 + \left(\mbf{q^\prime}_y-\mbf{r^\prime}_y\right)^2},\\
        \mbf{q^\prime} &\equiv \substack{\mathrm{argmin}\\\mbf{b^\prime}\in \overline{C^\prime}\\\mbf{b^\prime}_x \geq\mbf{p}_x}\left|\mbf{b^\prime}_x - \mbf{p}_x\right|,\\
        \mbf{r^\prime} &\equiv \substack{\mathrm{argmin}\\\mbf{b^\prime}\in \overline{C^\prime}\\\mbf{b^\prime}_x \leq \mbf{p}_x}\left|\mbf{b^\prime}_x - \mbf{p}_x\right|,
    \end{aligned}
\end{equation}
and $\Delta^2\left(\mbf{p}, \overline{C^\prime}\right)$ is the square of the uncertainty in that distance, found by propagating the uncertainties in the relevant parameters. Note that points which fall outside the domain of $C^\prime$ are given a loss of $0$, since they have no corresponding line segment in $C^\prime$. A diagram of this calculation is shown in \cref{fig:loss_example}.

From this formulation it is clear that the loss of a point in $\mbf{p}\in C$ that falls exactly on some part of $C^\prime$ is $0$, and that if the uncertainty of $C^\prime$ in the region of $\mbf{p}$ is large then the loss will be small.

We can then define the cost function between the two curves $C$ and $C^\prime$, $s\left(C, C^\prime\right)$ as the sum of the losses of each point in $\overline{C}$ relative to $\overline{C^\prime}$, and vice versa, normalized by the number of points which were not excluded in \cref{eq:loss},

\begin{equation}
    s\left(C, C^\prime\right) \equiv \frac{1}{2}\left[\frac{\sum\limits_{\mbf{p}\in\overline{C}} L\left(\mbf{p}, \overline{C^\prime}\right)}{\left|\overline{C}\right|_{\overline{C^\prime}}}+\frac{\sum\limits_{\mbf{p^\prime}\in\overline{C^\prime}} L\left(\mbf{p^\prime}, \overline{C}\right)}{\left|\overline{C^\prime}\right|_{\overline{C}}}\right],
\end{equation}
where

\begin{equation}
    \left|A\right|_B \equiv \sum\limits_{\substack{\mbf{a}\in A\\\min\limits_{\mbf{b}\in B}\mbf{b}_x \leq \mbf{a}_x \leq \max\limits_{\mbf{b}\in B}\mbf{b}_x}}
\end{equation}
denotes the number of elements in $A$ which fall within the domain of $B$.

Finally, the objective function for determining the collapse of a set of $N$ curves $C_1, C_2, \ldots, C_N$ is

\begin{equation}
    \label{eq:obj}
    S(C_1, C_2, \ldots, C_N) \equiv \frac{1}{N^2 - N}\sum\limits_{i<j}^N s\left(C_i, C_j\right).
\end{equation}
The normalization of $N^2 - N$ arises from the number of pairs of non-identical curves in the set.

\begin{figure}[H]
\centering
\subfloat[]{\includegraphics[width=0.45\linewidth]{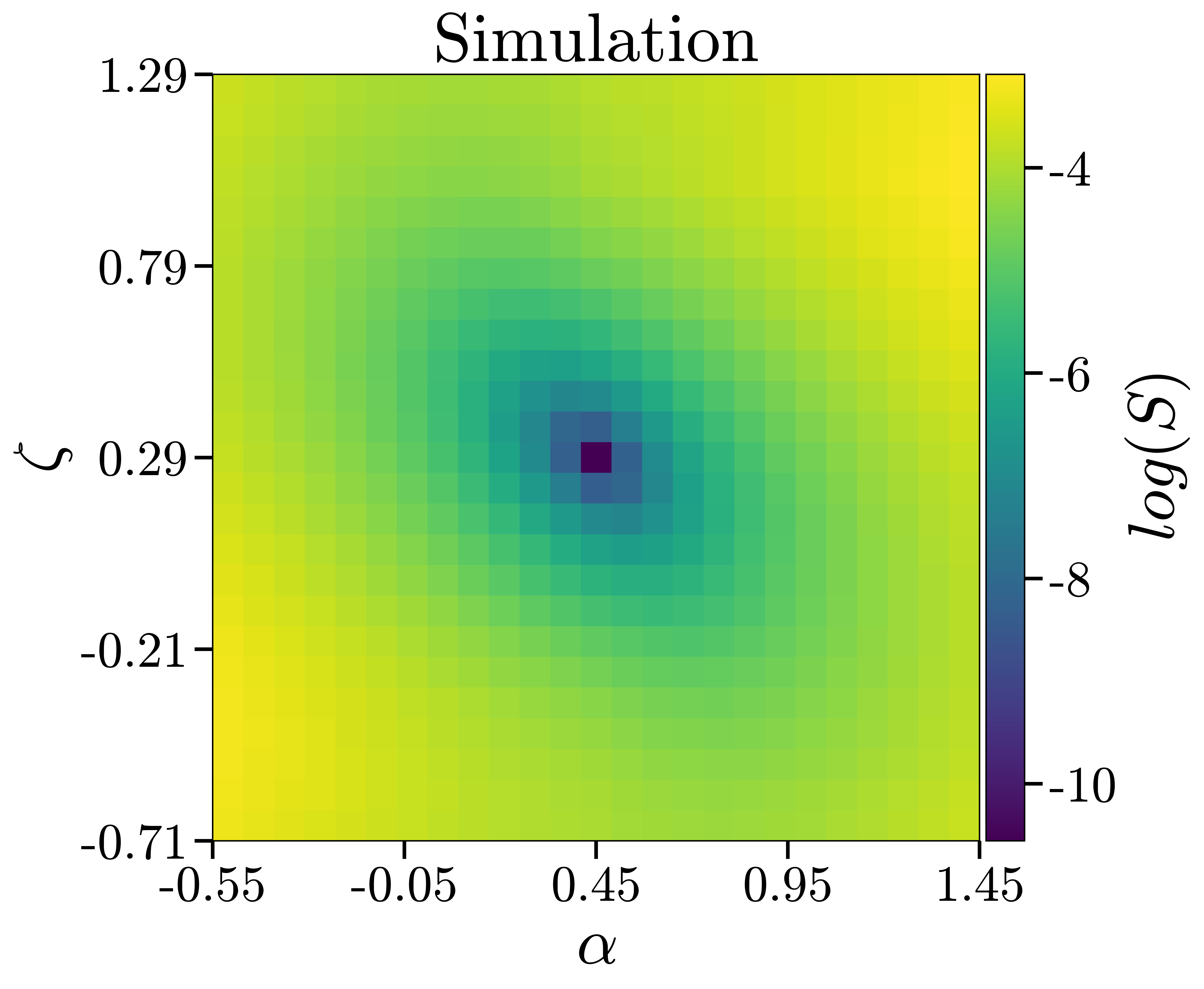}}\quad
\subfloat[]{\includegraphics[width=0.45\linewidth]{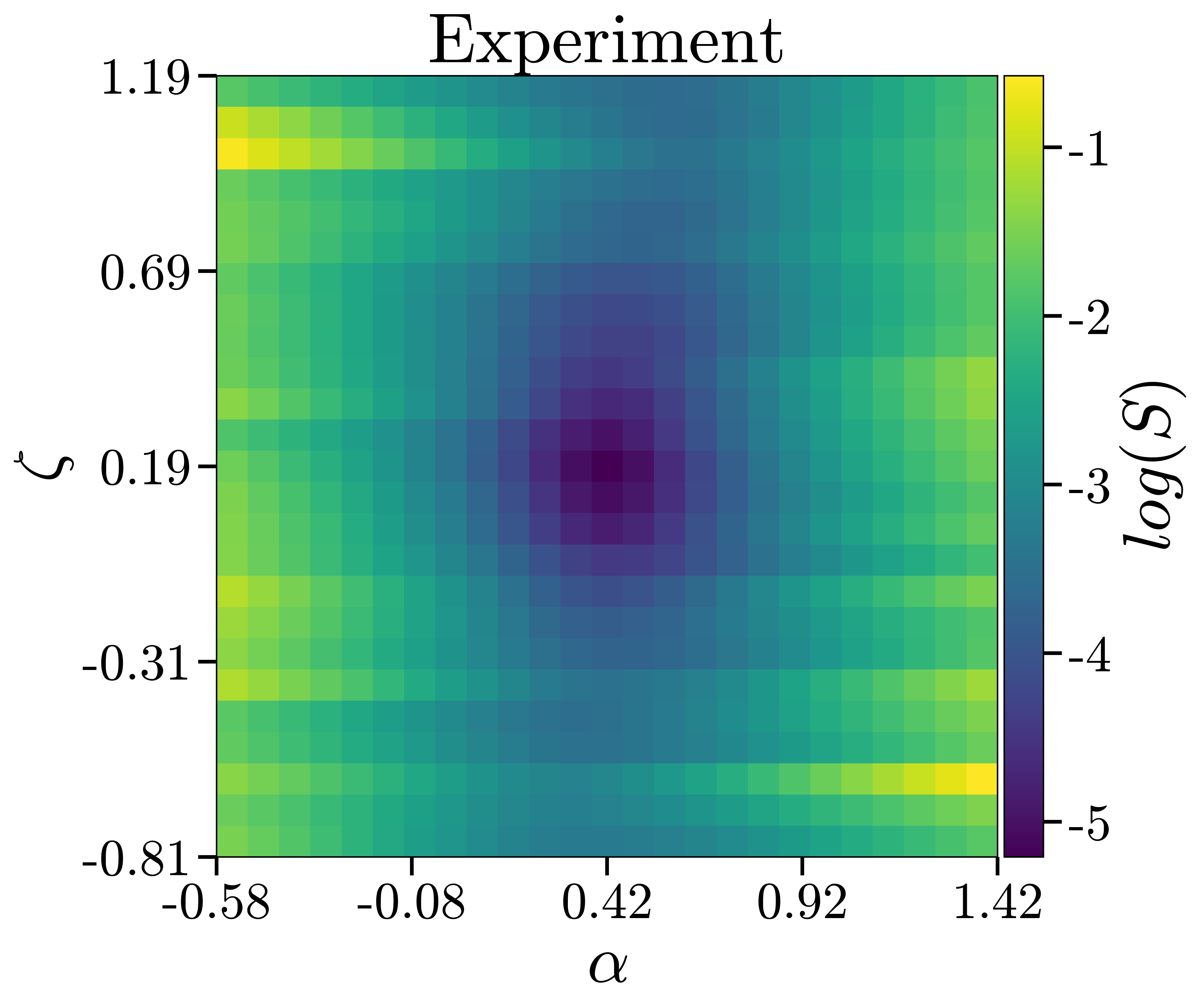}}\quad
\caption{\label{fig:S_manifold} Output space of the objective function, \cref{eq:obj}, around the calculated minimum for the (a) simulated and (b) experimental type-I quenches.}
\end{figure}

We perform convex optimization on this objective function with our data using Powell's method~\cite{Powell64} in order to determine the critical exponents which result in an optimal collapse. Supplementary~\cref{fig:S_manifold} shows the output space of our objective function, demonstrating that it is well-behaved and exhibits an identifiable minimum.

\subsection{Uncertainty Estimates via the Hessian}
\cref{eq:obj} incorporates the uncertainty in the data directly. To obtain an uncertainty in the predicted minimum location of the objective function, we use the Hessian. The uncertainty of any quantity $X$ as a function of the input parameters $\{x_i\}$ of an objective function $f$ can be estimated by~\cite{Pumplin01}

\begin{equation}
    \Delta\left(X\right)^2 = \left(\tilde{f} - \hat{f}\right) \sum\limits_{i,j} H_{ij}^{-1}\frac{\partial X}{\partial x_i}\frac{\partial X}{\partial x_j}.
\end{equation}
Where $H^{-1}$ is the inverse of the Hessian matrix of $f$ evaluated at the location of the predicted minimum, $\tilde{f}$ is the value of $f$ evaluated at the same location, and $\hat{f}$ is the true global minimum value of the objective function.

In our application $\{x_i\} = \{\alpha, \zeta\}$, and $X = \alpha$ or $\zeta$ as well. Moreover, we have designed our objective function such that the global minimum occurs when the curves perfectly collapse and so $\hat{f} = 0$. Therefore the uncertainty in the predicted exponents $\tilde{\alpha}$ and $\tilde{\zeta}$ is simply given by

\begin{equation}
    \begin{aligned}
        \Delta\tilde{\alpha} &= \sqrt{S\left(\tilde{\alpha}, \tilde{\zeta}\right)H_{\alpha\alpha}^{-1}},\\
        \Delta\tilde{\zeta} &= \sqrt{S\left(\tilde{\alpha}, \tilde{\zeta}\right)H_{\zeta\zeta}^{-1}}.
    \end{aligned}
\end{equation}

\section*{Supplementary References}